\documentclass[letterpaper, journal, twoside, 10pt,twocolumn]{support/IEEEtran}
\usepackage[fleqn]{amsmath}
\usepackage{times}
\usepackage[pdftex]{graphicx}
\usepackage{subfigure}
\usepackage{amsmath,amssymb,amsopn,amstext,amsfonts}
\usepackage{cancel}
\usepackage[noadjust]{cite}
\usepackage{soul}
\usepackage{caption}
\usepackage{balance}
\usepackage{color}
\usepackage{mathtools}
\usepackage{bm}
\usepackage{ diagbox}
\usepackage{float}
\usepackage{epstopdf}
\usepackage{url}
\usepackage{multirow}
\usepackage{tikz}
\usepackage{subeqnarray}
\usepackage{cases}
\usepackage{booktabs}
\usepackage[linkcolor=black,citecolor=black,urlcolor=black,colorlinks=true]{hyperref}
\usepackage{algorithm}
\usepackage[noend]{algpseudocode}
\newtheorem{myTheo}{Theorem}
\newtheorem{myDef}{Definition} %??????defn???thm????
\newtheorem{Lemma}{Lemma} %??????lem???thm????

\newtheorem{remark}{Remark}

\newtheorem{assumption}{Assumption}

\soulregister\cite7
\soulregister\citep7
\soulregister\citet7
\soulregister\ref7
\soulregister\it7
\soulregister\pageref7

\graphicspath{{figures/}}
\DeclareGraphicsExtensions{.pdf,.png,.jpg,.eps}
\IEEEoverridecommandlockouts
\title{\LARGE \bf Resilient Output Containment Control of Heterogeneous Multiagent Systems Against Composite Attacks: A Digital Twin Approach}

\author{
  \vskip 1em
  {Yukang Cui, \emph{Member, IEEE},
  Lingbo Cao,
  Michael V. Basin,  \emph{Senior Member, IEEE},
  Jun Shen,  \emph{Senior Member, IEEE},
  Tingwen Huang, \emph{Fellow, IEEE},
  Xin Gong, \emph{Member, IEEE}
  }

  \thanks{
    This work was partially supported by the National Natural Science Foundation of China under Grant 61903258, Guangdong Basic and Applied Basic Research Foundation 2022A1515010234 and the Project of Department of Education of Guangdong Province 2022KTSCX105. (\emph{Corresponding author: Xin Gong.}) %the National Natural Science Foundation of China under Grant 61903258

Y. Cui and L. Cao are with the College of Mechatronics and Control Engineering, Shenzhen University, Shenzhen, 518060, China (e-mail: {\tt\small cuiyukang,lingbcao@gmail.com}).

M. V. Basin is with School of Physical and Mathematical Sciences, the Autonomous University of Nuevo Leon, Mexico (e-mail:
{\tt\small mbasin@fcfm.uanl.mx}).

J. Shen is with College of Automation Engineering, Nanjing University of Aeronautics and Astronautics, Nanjing, 211106, China ({\tt\small junshen2009@gmail.com}).

T. Huang is with Texas A\&M University at Qatar, Doha, 23874, Qatar (e-mail: {\tt\small tingwen.huang@qatar.tamu.edu}).

X. Gong is with the Department of Mechanical Engineering, The University of Hong Kong, Pokfulam Road, Hong Kong (e-mail: {\tt\small gongxin@connect.hku.hk}).

  }
}

\begin{document}
  \maketitle
  \begin{abstract}
 This paper studies the distributed resilient output containment control of heterogeneous multiagent systems against composite attacks, including denial-of-services (DoS) attacks, false-data injection (FDI) attacks, camouflage attacks, and actuation attacks. Inspired by digital twins, a twin layer (TL) with higher security and privacy is used to decouple the above problem into two tasks: defense protocols against DoS attacks on TL and defense protocols against actuation attacks on cyber-physical layer (CPL). First, considering modeling errors of leader dynamics, we introduce distributed observers to reconstruct the leader dynamics for each follower on TL under DoS attacks. Second, distributed estimators are used to estimate follower states according to the reconstructed leader dynamics on the TL. Third, according to the reconstructed leader dynamics, we design decentralized solvers that calculate the output regulator equations on CPL. Fourth,
decentralized adaptive attack-resilient control schemes that resist unbounded actuation attacks are provided on CPL. Furthermore, we apply the above control protocols to prove that the followers can achieve uniformly ultimately bounded (UUB) convergence, and the upper bound of the UUB convergence is determined explicitly. Finally, two simulation examples are provided to show the effectiveness of the proposed control protocols.
\end{abstract}
\section{Introduction}
\IEEEPARstart{D}{ISTRIBUTED} cooperative control of multiagent systems (MASs) has attracted extensive attention over the last decade due to its broad applications in satellite formation \cite{scharnagl2019combining}, mobile multirobots \cite{wang2017adaptive}, and smart grids \cite{ansari2016multi}.
%  multi-robot cooperative control, ,
%  mobile multirobot [1], distributed microgrid [2], and vehicle platoon
% physics, biology, social activity, and engineering.
Consensus problems are one of the most common phenomena in the cooperative control of MASs and can be divided into two categories: leaderless consensus problem \cite{fax2004information,ren2007information,olfati2007consensus,deng2020dynamic} and leader-follower consensus ones \cite{hong2013distributed,zhao2016distributed,su2011cooperative,su2012cooperative}.
%\cite{,,kim2010output,wang2010distributed,lunze2012synchronization,yaghmaie2016output}
Those control schemes drive all agents to reach an agreement on certain variables of interest. In recent years, researchers have introduced multiple leaders into consensus problems, and the control schemes are developed to drive certain variables of all followers into the convex hulls formed by the leaders. This phenomenon is also known as the containment control problem \cite{haghshenas2015,zuo2017output,zuo2019,zuo2020,lu2021,ma2021observer}. Haghshenas et al. \cite{haghshenas2015} and Zuo et al. \cite{zuo2017output} solved the containment control problem of heterogeneous linear MASs with output regulator equations. Then, Zuo et al. \cite{zuo2019} considered the containment control problem of heterogeneous linear MASs against camouflage attacks. Moreover, Zuo et al. \cite{zuo2020} presented an adaptive compensator against unknown actuator attacks. Lu et al. \cite{lu2021} considered containment control problems against false-data injection (FDI) attacks. Ma et al. \cite{ma2021observer} solved the containment control problem against denial-of-service (DoS) attacks by developing event-triggered observers. However, the above containment control works only for a single type of attacks.
Considering the vulnerability of MASs, the situation may be more complicated in reality, that is, the systems may face multiple attacks at the same time. Therefore, this paper considers the multiagent resilient output containment control problem against composite attacks, including DoS attacks, FDI attacks, camouflage attacks, and actuation attacks.

In recent years, digital twins \cite{zhu2020,gehrmann2019digital} have been actively used in various fields, including industry \cite{abburu2020cognitwin}, aerospace
%\cite{salinger2020hardware}
\cite{liu2022intelligent}, and medicine \cite{voigt2021digital}. The digital twin technology involves the real-time mapping of objects in the virtual space. Inspired by this technology, in this paper, a double-layer control structure with a twin layer (TL) and a cyber-physical layer (CPL) is designed. The TL is a virtual mapping of the CPL that has the same communication topology as the CPL and can interact with the CPL in real time. Moreover, the TL has high privacy and security and can effectively address data manipulation attacks in communication networks, such as FDI attacks and camouflage attacks. Therefore, the main idea of this paper is to design a corresponding control framework that resists DoS attacks
and actuation attacks. Since the TL is only affected by DoS attacks, this paper can be divided into two parts: one is to resist DoS attacks on the TL and the other is to resist actuation attacks on the CPL.

Resilient control protocols for MASs against DoS attacks were studied in \cite{fengz2017,zhangd2019,zong2022,yang2020,deng2021}. Feng et al. \cite{fengz2017} considered the leader-follower consensus through its latest state and leader dynamics during  DoS attacks. Yang et al. \cite{yang2020} designed a dual-terminal event-triggered mechanism against DoS attacks for linear leader-following MASs.
On the other hand, the methods proposed in the above paper assume that the leader dynamics is known to each follower, which is unrealistic in some cases. To address this issue, Cai et al. \cite{cai2017} and Chen et al. \cite{chen2019} used adaptive distributed observers to estimate the leader dynamics and states for an MAS with a single leader. Deng et al. \cite{deng2021} proposed a distributed resilient learning algorithm for unknown dynamic models. However, those papers do not consider leader estimation errors and DoS attacks simultaneously. Thus, motivated by the above results, we consider a TL with modeling errors of the leader dynamics, which is inevitable in online modeling. This is a challenging problem, since modeling errors of leader dynamics affect follower state estimation errors and output containment errors. To address this problem, we introduce a distributed TL that handles modeling errors of leader dynamics and DoS attacks. On this TL, distributed observers are designed to reconstruct the leader dynamics of each follower. Then, distributed state estimators are used to reconstruct follower states under DoS attacks.

For the defense strategy against actuation attacks, a pioneering paper \cite{de2014resilient} presented a distributed resilient adaptive control scheme against attacks with exogenous disturbances and inter-agent uncertainties. Xie et al. \cite{xie2017decentralized} considered distributed adaptive fault-tolerant control problem for uncertain large-scale interconnected systems with external disturbances and actuator attacks. Moreover, Jin et al. \cite{jin2017adaptive} provided an adaptive controller guaranteeing that closed-loop dynamics against time-varying adversarial sensor and actuator attacks can achieve uniformly ultimately bounded (UUB) performance. However, the above paper consider only bounded attacks (or disturbances), which cannot maximize their destructive capacity indefinitely. Recently, Zuo et al. \cite{zuo2020} provided an adaptive control scheme for MASs against unbounded actuator attacks.
Based on the above results, we present a decentralized CPL to address the output containment problem against unbounded actuation attacks. On this CPL, we solve the output regulator equation with reconstructed leader dynamics to address the output containment problem for heterogeneous MASs. Then, decentralized attack-resilient protocols with adaptive compensation signals are used to resist unbounded actuation attacks.

%Moreover, in contrast to \cite{zuo2020}, our adaptive compensation signal includes a soft-sign function, and the containment error bound can be determined explicitly.

%Editor: Please ensure that the intended meaning has been maintained in the following edit.
The above discussion indicates that the output containment control problem for heterogeneous MASs with TL modeling errors of leader dynamics under the composite attacks has  not yet been solved or studied.
Therefore, in this paper, we provide a hierarchical control scheme to address the above problem.
Our main contributions can be summarized as follows:
\begin{enumerate}
  \item In contrast to previous results, our paper considers TL modeling errors of leader dynamics. Thus, the TL needs to consider both modeling errors  and DoS attacks, and the CPL needs to consider both output regulator equation errors and actuation attacks, thereby increasing the complexity and practicality of the analysis. In addition, compared with \cite{deng2021}, where the state estimator requires some time to estimate the leader dynamics, in this paper, the leader dynamic observers and follower state estimators on the TL work synchronously.
  \item A double layer control structure is introduced, and a TL is established in the control framework. The TL has higher security than the CPL, that is, it can effectively defend against various types of attacks, including FDI attacks and camouflage attacks. Moreover, due to the existence of the TL, the resilient control scheme can be decoupled into defense against DoS attacks on the TL and defense against potentially unbounded actuation attacks on the CPL.
  \item Compared with \cite{zuo2020}, our paper considers composite attacks and employs an adaptive compensation input against unbounded actuation attacks, which yields UUB convergence. The containment output error bound of the UUB performance is given explicitly.
\end{enumerate}

\noindent\textbf{Notations:}
In this paper, $I_r$ is the identity matrix with compatible dimensions.
$\boldsymbol{1}_m$ (or $\boldsymbol{0}_m$) denotes a column vector of size $m$ with values of $1$ (or $0$, respectively). Denote the index set of sequential integers as $\textbf{I}[m,n]=\{m,m+1,\ldots~,n\}$, where $m<n$ are two natural numbers. Define the set of real numbers and the set of natural numbers as $\mathbb{R}$ and $\mathbb{N}$, respectively. For any matrix $A\in \mathbb{R}^{M\times N}$, ${\rm vec }(A)= {\rm col}(A_1,A_2,\dots,A_N)$, where $A_i\in \mathbb{R}^{M} $ is the $i$th column of $A$.
Define ${\rm {\rm blkdiag}}(A_1,A_2,\dots,A_N)$ as a block diagonal matrix whose principal diagonal elements are equal to the given matrices $A_1, A_2, \dots, A_N$. $\sigma_{\rm min}(A)$, $\sigma_{\rm max}(A)$ and $\sigma(A)$ denote the minimum singular value, maximum singular value and spectrum of matrix $A$, respectively. $\lambda_{\rm min}(A)$ and $\lambda_{\rm max}(A)$ represent the minimum and maximum eigenvalues of $A$, respectively. Herein, $||\cdot||$ denotes the Euclidean norm, and $\otimes$ denotes the Kronecker product.
%; moreover, $A>0$ means that $\lambda_1(A)>0$.
%For a time-varying function $x(t): \mathbb{R}_{\geq 0 }\mapsto \mathbb{R}$, denote that $\sup_{t\in [t_0, t_1]} x(t) $ and $\inf_{t\in [t_0, t_1]} x(t)$ as the upper bound and lower bound of $x(t)$ over the time interval $[t_0, t_1]$, respectively. Moreover, denote that $\|x(t)\|_{[t_0, t_1]} =\sup_{t\in [t_0, t_1]} \|x(t)\| $. Define that $L_{\infty}:=\{x(t)|x(t): \mathbb{R}_{\geq 0 }\mapsto \mathbb{R}^n,\ \|x(t)\|_{[t_0, t_1]}<\infty\}$. In the following sections, $x(t) \in L_{\infty}$, $t\in [t_0, t_1]$, represents that the variable $x$ is uniformly bounded over $[t_0, t_1]$.   %$A>eq 0$ (or $A> 0$) denotes that $A$ is a nonnegative matrix (positive matrix, respectively), which means all elements of $A$ are nonnegative (positive, respectively).
 %${\rm span}(x)$ denotes the span vector of a given vector $x=[p_1, p_2,\ldots~, p_n]^{\mathrm{T}}\in \mathbb{R}^n$.

\label{introduction}

\section{Preliminaries}\label{section2}

%\subsection{Notations}

{\color{black}
\subsection{Graph Theory}
A directed graph is used to represent interactions among agents. For an MAS with $n$ agents, the graph
$\mathcal{G}$ can be expressed by $(\mathcal{V}, \mathcal{E}, \mathcal{A} )$, where $\mathcal{V}=\{1, 2, \ldots~, N \}$ is the node set,
$\mathcal{E} \subset \mathcal{V} \times \mathcal{V}=\{(v_j,\ v_i)\mid v_i,\ v_j \in \mathcal{V}\}$ is the edge set, and $\mathcal{A}=[a_{ij}] \in \mathbb{R}^{N\times N} $ is the associated adjacency matrix.
The weight of edge $(v_j,\ v_i)$ is denoted by $a_{ij}$. if $(v_j, \ v_i) \in \mathcal{E}$, $a_{ij} > 0$,  and otherwise, $a_{ij} = 0$ . The neighbor set of node $v_i$ is represented by $\mathcal{N}_{i}=\{v_{j}\in \mathcal{V}\mid (v_j,\ v_i)\in \mathcal{E} \}$. Define the in-degree matrix as $\mathcal{D}={\rm blkdiag}(d_i) \in \mathbb{R}^{N\times N}$ with $d_i=\sum_{j \in \mathcal{N}_{i}} a_{ij}$.
The Laplacian matrix is denoted as
$L=\mathcal{D}-\mathcal{A}  \in \mathbb{R}^{N\times N}$.
%A directed graph is \emph{strongly connected} if there is a path from $v_i$ to $v_j$ for any pair of nodes $(v_i,v_j)$.
}
\subsection{Lemmas and Definitions}
\begin{myDef}[{\rm \cite{khalil2002nonlinear}}]
The signal $x(t)$ is UUB with the ultimate bound $b$ if there exist $b>0$ and $c>0$ that are independent of $t_0 \geq 0$ and for every $a \in (0, c)$ if there exists $T = T(a, b) \geq 0$ , independent of $t_0$, such that
 \begin{equation}
     ||x(t)||\leq a \Rightarrow ||x(t)|| \leq b,~\forall t \in [t_0+T,\infty).
 \end{equation}
\end{myDef}
\begin{Lemma}[{\rm Bellman-Gronwall Lemma \cite{lewis2003}}] \label{Bellman-Gronwall Lemma 2}
    Assume that $\Phi : [T_a,T_b] \rightarrow \mathbb{R}$ is a nonnegative continuous function,  $\alpha: [T_a, T_b] \rightarrow \mathbb{R}$ is a continuous function, $\kappa \geq 0$ is a constant, and
    \begin{equation}
    \Phi(t) \leq \kappa + \int_{0}^{t} \alpha (\tau)\Phi(\tau) \,{\rm d}\tau,~t \in [T_a,T_b];
    \end{equation}
    then, we have
    $$\Phi(t) \leq \kappa \mathrm{e}^{\int_{0}^{t} \alpha(\tau)\,{\rm d}\tau }$$ for all $t \in [T_a, T_b]$.
\end{Lemma}
\section{System Setup and Problem Statement}\label{section3}
In this section, a new problem, namely, the resilient containment of
MAS groups against composite attacks, is considered. First, a model of the MAS group is established and some composite attacks are defined.
{\color{black}
\subsection{MAS Group Model}
In the framework of containment control, we consider a group of $N+M$ MASs, which can be divided into two subgroups:

1) The $M$ leaders are the roots of the directed graph $\mathcal{G}$ and have no neighbors. We define the index set of the leaders as $\mathcal{L}= \textbf{I}[N+1,N+M]$.

2) The $N$ followers coordinate with their neighbors to achieve the containment set
of the above leaders. We define the index set of the followers
as $\mathcal{F} = \textbf{I}[1, N]$.

Similarly to the previous works \cite{zuo2020,zuo2021,haghshenas2015}, we consider the following leader dynamics:
\begin{equation}\label{EQ1}
\begin{cases}
\dot{x}_k=S x_k,\\
y_k=R x_k,
\end{cases}
\end{equation}
where $x_k\in \mathbb{R}^q$ is $k$th leader state, $y_k\in \mathbb{R}^p$ is the $k$th leader reference output,  with $k \in \mathcal{L}$.
The dynamics of each follower is given by
\begin{equation}\label{EQ2}
  \begin{cases}
  \dot{x}_i=A_i x_i + B_i u_i,\\
  y_i=C_i x_i,
  \end{cases}
\end{equation}
where $x_i\in \mathbb{R}^{ni}$, $u_i\in \mathbb{R}^{mi}$, and $y_i\in \mathbb{R}^p$ are
the system state, the control input, and the output of the $i$th follower, with $i\in \mathcal{F}$, respectively. The leader $(S,R)$ and follower $(A_i,B_i,C_i)$ dynamics may have different dimensions.    For convenience, the notation $``(t)"$ can be omitted in the following discussion. We make the following assumptions about the agents and communication network.

{\color{black}
\begin{assumption}\label{assumption 1}
There exists a directed path from at least one leader to each follower $i\in\mathcal{F}$ in graph $\mathcal{G}$.
\end{assumption}

\begin{assumption}\label{assumption 2}
  The real parts of the eigenvalues of $S$ are nonnegative.
\end{assumption}

\begin{assumption}\label{assumption 3}
The pair $(A_i, B_i)$ is stabilizable and the pair $(A_i, C_i)$ is detectable for $i \in \mathcal{F}$.
\end{assumption}

\begin{assumption}\label{assumption 4}
For all $\lambda \in \sigma(S)$, where $\sigma(S)$ represents the spectrum of $S$, we have
  \begin{equation}
   {\rm rank} \left[
      \begin{array}{c|c}
     A_i-\lambda I_{n_i} &  B_i  \\ \hline
    C_i  & 0   \\
      \end{array}
      \right]=n_i+p,~
      i \in \mathcal{F}.
  \end{equation}
\end{assumption}

}

\begin{remark}
Assumptions 1, 3 and 4 are standard in conventional output regulation problems, and Assumption 2 prevents considering the trivial case of a stable $S$. The modes associated with eigenvalues of $S$ with negative real parts decay exponentially to zero and therefore do not affect the asymptotic behavior of the closed-loop system. $\hfill \hfill \square $
\end{remark}

}

\subsection{Attack Descriptions}
In this paper, we consider an MAS consisting of cooperative agents with potential malicious attackers. As shown in Fig. \ref{fig:figure0}, the attackers use four kinds of attacks to compromise the containment performance of the MAS:

\begin{figure}[!]
  %\begin{minipage}[t]{1\linewidth}
  \centering
  \includegraphics[width=0.45\textwidth]{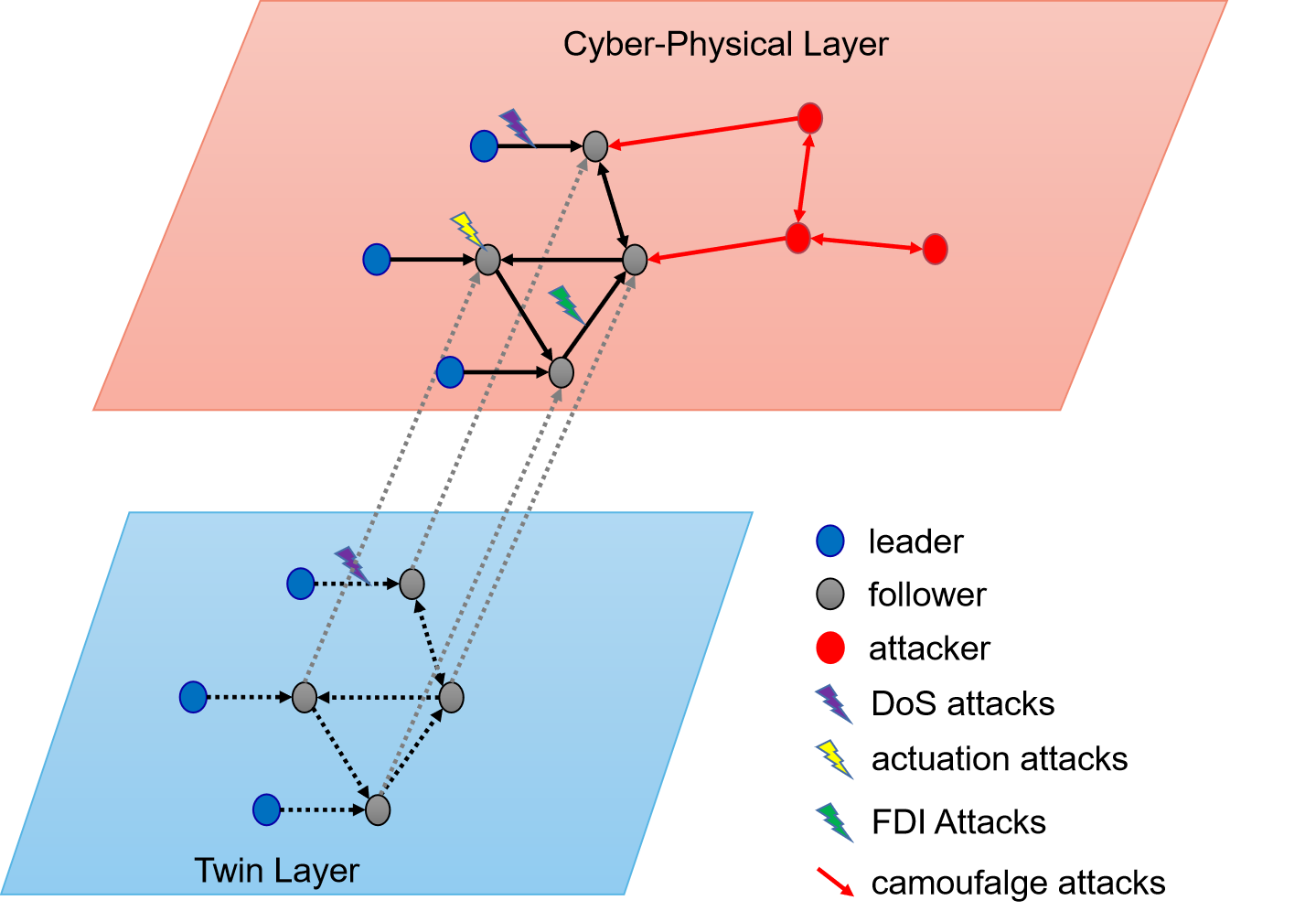}
  \caption{Resilient MASs against composite attacks: A double layer framework.}
  \label{fig:figure0}
\end{figure}

1) DoS attacks: The communication graphs among the agents (in both the TL and CPL) are destroyed by the attackers;

2) Actuation attacks: The motor inputs are infiltrated by the attackers to distort the input signals of the agent;

3) FDI attacks: The information exchanged among the agents is distorted by the attackers;

4) Camouflage attacks: The attackers mislead downstream agents by disguising themselves as leaders.

To resist composite attacks, we introduce a new layer known as TL. The TL has the same communication topology as the CPL with greater security and fewer physical meanings. Therefore, the TL can effectively resist most of the above attacks. With the introduction of the TL, a decoupled resilient control scheme can be introduced to defend against DoS attacks on the TL and potential unbounded actuation attacks on the CPL. The following subsections present the definitions and essential constraints of the DoS and actuation attacks.

1) DoS attacks: DoS attacks refer to attacks where an adversary attacks some or all components of a control system. DoS attacks affect the measurement and control channels simultaneously, resulting in the loss of data availability. Define $\{t_l \}_{l \in \mathbb{N}}$ and $\{T_l \}_{l \in \mathbb{N}}$ as the start and terminal time of the $l$th attack sequence of a DoS attack, that is, the $l$th DoS attack time interval is $A_l = [t_l, T_l)$, where $t_{l+1} > T_l$ for all $l \in \mathbb{N}$. Therefore, for all $t\geq t_0 \in \mathbb{R}$, the set of time instants where the communication network is under DoS attacks can be represented by
\begin{equation}
    \Xi_A(t_0,t) = \cup A_l \cap [t_0, t],~l\in \mathbb{N}.
\end{equation}
Moreover, the set of time instants of the destroyed communication network can be defined as
\begin{equation}
    \Xi_N(t_0,t) = [t_0,t] / \Xi_A(t_0,t).
\end{equation}

\begin{myDef} [{\rm Attack frequency \cite{fengz2017}}]
For any $\delta_2 > \delta_1 \geq t_0$, let $N_a(t_1, t_2)$ represent the number of DoS attacks during $[t_1, t_2)$. Therefore, $F_a(t_1, t_2)= \frac{N_a(t_1, t_2)}{t_2 - t_1}$ is defined as the attack frequency during $[t_1, t_2)$ for all $t_2 > t_1 \geq t_0$.
\end{myDef}

\begin{myDef} [{\rm Attack duration \cite{fengz2017}}]\label{Def3}
For any $t_2 > t_1 \geq t_0$, let $T_a(t_1, t_2)$ represent the total time interval of DoS attacks on MASs during $[t_1, t_2)$. The attack duration over $[t_1, t_2)$ can be defined with constants $\tau_a > 1$ and $T_0 > 0$ as $T_a(t_1,t_2) \leq T_0 + \frac{t_2-t_1}{\tau_a}$.
\end{myDef}

2) Unbounded Actuation Attacks:
Under the influence of actuation attacks, the control input of each follower is defined as
\begin{equation}\label{EQ9}
    \bar{u}_i=u_i+\chi_i,~ \forall i  \in \mathcal{F},
\end{equation}
where $\chi_i$ denotes an unknown unbounded attack signal. Thus, the true values of $u_i$ and $\chi_i$ are unknown, and we can only measure the damaged control input information $\bar{u}_i$.

\begin{assumption}\label{assumption 5}
The unknown actuation attack signal $\chi_i$ is unbounded, but its derivative $\dot{\chi}_i$ is bounded by $\bar{d}$.
\end{assumption}

{\color{black}
\begin{remark}
In contrast to \cite{deng2021} and \cite{chen2019}, which only considered bounded actuator attacks, this paper deals with composite attacks, including unbounded actuation attacks under Assumption \ref{assumption 5}. When the derivative of the attack signal is unbounded, the attack signal increases substantially, and the MAS can reject the signal by removing excessively large derivative values, which can easily be detected.
$\hfill \hfill \square$
\end{remark}

}

\subsection{Problem Statement}

%{\color{red}
Under the condition that there exists no attack, the output containment error of the $i$th follower can be represented as
\begin{equation}\label{EQ xi}
    \xi_i = \sum_{j\in \mathcal{F}} a_{ij}(y_j -y_i) +\sum_{k \in \mathcal{L}} g_{ik}(y_k - y_i),
\end{equation}
where $a_{ij}$ is the weight of edge $(v_i, v_j)$ in graph $\mathcal{G}_f$
($\mathcal{G}_f$ is the communication graph of all followers),
and $g_{ik}$ is the weight of the path between the $i$th leader and the $k$th follower.

The global form of (\ref{EQ xi}) is represented as
\begin{equation}
    \xi = - \sum_{k \in \mathcal{L}}(\Psi_k \otimes I_p)(y -  \underline{y}_k),
\end{equation}
where $\Psi_k = (\frac{1}{m} L_f + G_{ik})$, $L_f$ is the Laplacian matrix of communication digraph $\mathcal{G}_f$, and $G_{ik}={\rm diag}(g_{ik})$, $\xi = [\xi_1^{\mathrm{ T}},\xi_2^{\mathrm{T}},\dots,\xi_n^{\mathrm{T}}]^{\mathrm{T}}$, $y=[y_1^{\mathrm{T}},y_2^{\mathrm{T}},\dots,y_n^{\mathrm{T}}]^{\mathrm{T}}$, $\underline{y}_k = (l_n \otimes y_k)$.

\begin{Lemma}[{\rm \cite{haghshenas2015}}]\label{Lemma3}
   Suppose that Assumption 1 holds. Then, matrices $\Psi_k$ and $\bar{\Psi}_L= \sum_{k \in \mathcal{L}} \Psi_k$ are positive-definite and nonsingular. Therefore, $(\Psi_k)^{-1}$ and $(\bar{\Psi}_L)^{-1}= (\sum_{k \in \mathcal{L}} \Psi_k)^{-1}$ are both nonsingular.
\end{Lemma}

The global output containment error can be written as \cite{zuo2019}:
\begin{equation}\label{EQ13}
e= y - \left(\sum_{r\in \mathcal{L} }(\Psi_r \otimes I_p)\right)^{-1} \sum_{k \in \mathcal{L} } (\Psi_k \otimes I_p) \underline{y}_k,
\end{equation}
where $e=[e_i^{\mathrm{T}},e_2^{\mathrm{T}},\dots,e_n^{\mathrm{T}}]^{\mathrm{T}}$ and $\xi = -\sum_{k \in \mathcal{L}}(\Psi_k \otimes I_p )e$.

\begin{Lemma}
    Define $\Theta={\rm diag}(v)$, where $v=(\bar{\Psi}_L)^{-1} \boldsymbol{1}_N$. Under Assumption 1 and Lemma \ref{Lemma3}, $\Omega=\Theta \bar{\Psi}_L +\bar{\Psi}_L^{\mathrm{T}} \Theta >0$ is a positive definite diagonal matrix.
\end{Lemma}

\begin{myDef}\label{def41}
  For the $i$th follower, the system achieves containment if there exists a series of $\alpha_{\cdot i}$ that satisfy $\sum _{k \in \mathcal{L}} \alpha_{k i} =1$, thereby ensuring that the following equation holds:
   \begin{equation}
     {\rm  lim}_{t\rightarrow \infty } \left(y_i(t)-\sum _{k\in \mathcal{L}} \alpha_{k i}y_k(t)\right)=0,
   \end{equation}
   where $i \in \textbf{I}[1, N]$.
\end{myDef}

\begin{Lemma}[\rm {\cite[Lemma 1]{zuo2019}}]
    According to Assumption 1, the output containment control objective in Definition \ref{def41} is achieved if ${\rm lim}_{t \rightarrow \infty} e = 0$.
\end{Lemma}

Considering the composite attacks discussed in Subsection III-B, the local neighboring relative output containment information can be rewritten as
\begin{equation}\label{EQ xi2}
    \bar{\xi}_i = \sum_{j\in \mathcal{F}} d_{ij}(\bar{y}_{i,j} -\bar{y}_{i,i}) +\sum_{k \in \mathcal{L}} d_{ik}(\bar{y}_{i,k} - \bar{y}_{i,i})+\sum_{l\in {\mathcal{C}}} c_{il}(y_l-\bar{y}_{i,i}),
\end{equation}
where $d_{ij}(t)$ and $d_{ik}(t)$ are edge weights influenced by the DoS attacks. For denied communication, $d_{ij}(t)=0$ and $d_{ik}(t)=0$; for normal communication, $d_{ij}(t)=a_{ij}$ and $d_{ik}(t)=g_{ik}$. In addition, $\bar{y}_{i,j}$ (or $\bar{y}_{i,i}$) represents the information flow between $y_j$ and the $i$th follower compromised by FDI attacks. $\mathcal{C}$ is the index set of all camouflage attackers. $c_{il}$ represents the edge weight of the $l$th camouflage attack on the $i$th follower. According to (\ref{EQ xi2}), attackers can disrupt communication between agents and distort the actuation inputs of all followers.

Based on the above discussion, the attack-resilient containment control problem is introduced as follows.
%Next, we introduce the attack-resilient containment control problem.

\noindent \textbf{Problem ACMCA} (Attack-resilient Containment control of MASs
against Composite Attacks): The resilient containment control problem consists in designing the input $u_i$ in (\ref{EQ2}) for each follower such that the output containment error $e$ in (\ref{EQ13}) is UUB under composite attacks, i.e., the trajectories of each follower converge to a vicinity of point in the dynamic convex hull spanned by the trajectories of multiple leaders.

%}

\section{Main Results}

\begin{figure}[!]
  %\begin{minipage}[t]{1\linewidth}
  \centering
  \includegraphics[width=0.45\textwidth]{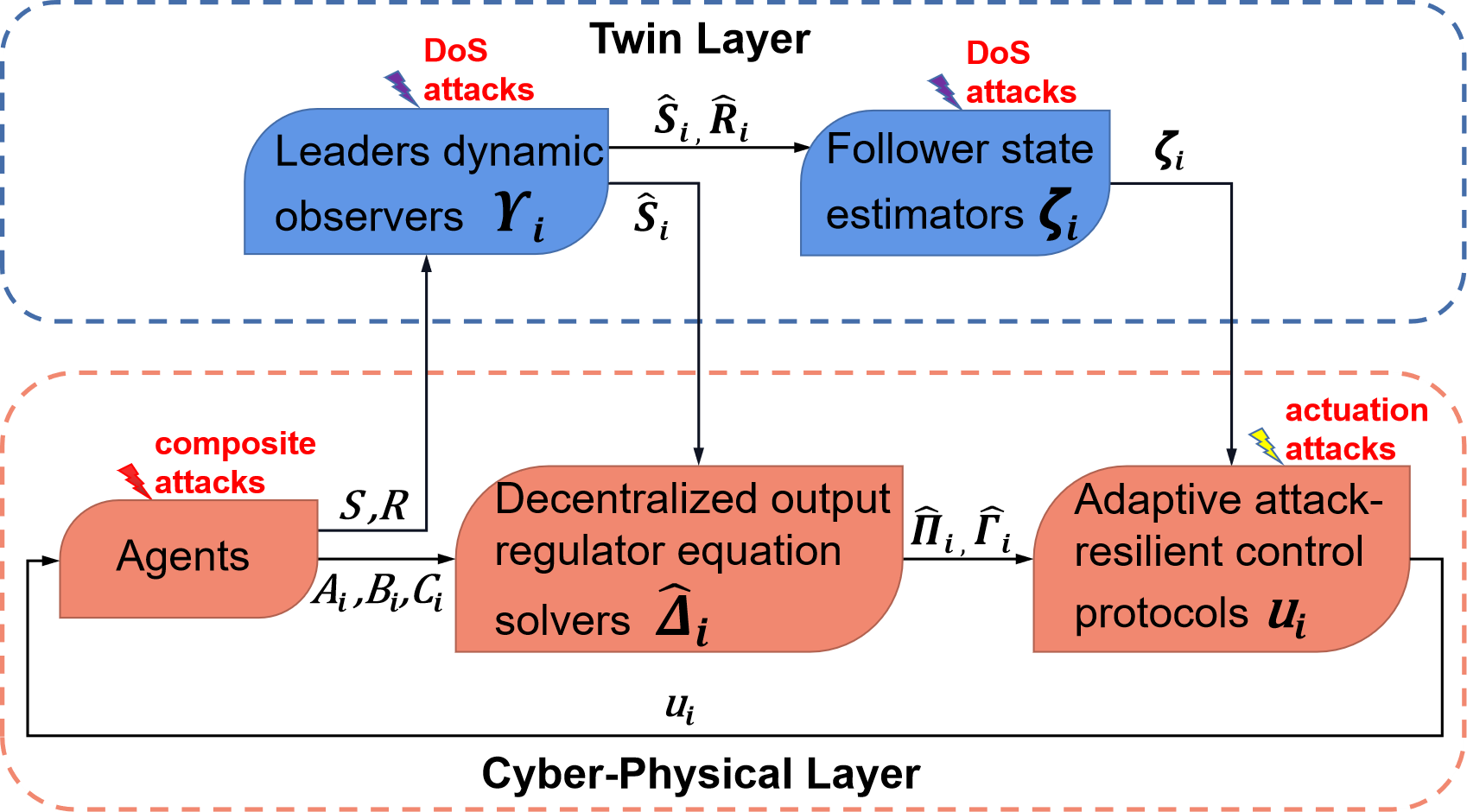}
  \caption{The ACMCA control framework.}
  \label{fig:figure0c}
\end{figure}

In this section, a double layer resilient control scheme is used to solve the \textbf{Problem ACMCA}. First, a distributed virtual resilient TL that can resist FDI attacks, camouflage attacks, and actuation attacks is proposed to cope with DoS attacks. Considering TL modeling errors of leader dynamics, we use distributed observers to reconstruct the leader dynamics of all agents under DoS attacks. Next, distributed estimators are proposed to reconstruct the follower states under DoS attacks. Then, we use the reconstructed leader dynamics proposed in Theorem 1 to solve the output regulator equation. Finally, adaptive controllers are proposed to resist unbounded actuation attacks on the CPL. The specific control framework is shown in Fig. \ref{fig:figure0c}.

\subsection{Distributed Observers for Reconstructing Leader Dynamics}
In this section, we design distributed observers to reconstruct the leader dynamics on the TL.

 To facilitate the analysis, we rewrite the leader dynamics as
  \begin{equation}
      \Upsilon =\left[\begin{array}{cc}
           S  \\
           R
      \end{array}\right]\in \mathbb{R} ^{(p+q)\times q},
  \end{equation}
and the reconstructed leader dynamics is introduced as
\begin{align} \label{EQ14}
     &\hat{\Upsilon } _{i}(t)=\left[\begin{array}{cc}
           \hat{S}_{i}(t)  \\
           \hat{R}_{i}(t) \\
      \end{array}\right] \in \mathbb{R} ^{(p+q)\times q},
  \end{align}
where $\hat{\Upsilon } _{i}(t)$ converges to $\Upsilon$ exponentially  by Theorem 1.

\begin{myTheo}\label{Theorem 1}
    Consider a group of $M$ leaders and $N$ followers composed of (\ref{EQ1}) and (\ref{EQ2}). Suppose that Assumption \ref{assumption 1} holds. Let the reconstructed leader dynamics $\hat{\Upsilon } _{i}(t)$, $i=\textbf{I}[1,N]$, in (\ref{EQ14}) on the TL be updated as
\begin{equation}\label{EQ15}
   \dot{\hat{\Upsilon }} _{i}(t)=  \mu_1 \! \left(\sum_{j \in \mathcal{F}} \! d_{ij} \left(\hat{\Upsilon } _{j}(t) \!-\! \hat{\Upsilon } _{i}(t) \!\right)
   \! +\! \sum_{k  \in \mathcal L} d_{ik} \! \left(\Upsilon\!-\!\hat{\Upsilon } _{i}(t) \!\right) \! \right),
\end{equation}
where $\mu_1 > \frac{\sigma_{\rm max}(S)}{ \lambda_{\rm min}(\Omega \Theta^{-1} ) (1-\frac{1}{\tau_a})}$ is the estimator gain. Then, the reconstructed leader dynamics $\hat{\Upsilon } _{i}(t)$ converge to $\Upsilon$ exponentially.
\end{myTheo}

%{\color{blue}
\textbf{Proof.}
From (\ref{EQ15}), it can be seen that we only use the relative neighborhood information to reconstruct the leader dynamics. Therefore, the leader dynamics observers are influenced by DoS attacks.

%\textbf{Step 1:}
{
Let
$\tilde{\Upsilon}_{i}(t)=\left[\begin{array}{cc}
           \tilde{S}_{i}(t)  \\
           \tilde{R}_{i}(t) \\
      \end{array}\right]=\left[\begin{array}{cc}
           \hat{S}_{i}(t) -S \\
           \hat{R}_{i}(t) -R\\
      \end{array}\right]$ denote the leader dynamic modeling error. According to (\ref{EQ15}),} we have
\begin{equation}\label{EQ17}
   \begin{aligned}
    \dot{\tilde{\Upsilon}} _{i} (t)
=&\dot{\hat{\Upsilon}}_{i} (t) -\dot{\Upsilon} \\
=&\mu_1\! \left( \sum_{j = 1}^{N}\! d_{ij} \! \left(\hat{\Upsilon } _{j}\!(t)\!-\!\hat{\Upsilon } _{i}\!(t) \! \right) \!+\!\sum_{k=N+1}^{N+M} \! d_{ik} \!\left(\Upsilon \!- \!\hat{\Upsilon } _{i}\!(t) \!\right)\! \right).
\end{aligned}
\end{equation}

Then, the global modeling error of the leader dynamics can be defined as
\begin{align}\label{EQ 18}
    \dot{\tilde{\Upsilon}}  (t) =& -  \mu_1  \left(\sum_{k\in \mathcal{L}}\Psi_k^D \otimes I_{p+q}\tilde{\Upsilon} (t)\right)\nonumber\\
    =&-  \mu_1 \left(\bar{\Psi}_{L}^D \otimes I_{p+q}\right) \tilde{\Upsilon} (t), ~t \geq t_0,
\end{align}
where $\bar{\Psi}^D_{L}=\sum_{k\in \mathcal{L}}\Psi_k^D $ with $\Psi_k^D (t) = \begin{cases}
0,~t \in \Xi_A, \\ \Psi_k,~t \in \Xi_N,
\end{cases}$ and $\tilde{\Upsilon} (t)= [\tilde{\Upsilon}_{1}(t)^{\mathrm{T}},\tilde{\Upsilon}_{1}(t)^{\mathrm{T}},\dots,\tilde{\Upsilon}_{N}(t)^{\mathrm{T}}]^{\mathrm{T}}$.
From (\ref{EQ 18}), we have ${\rm vec }(\dot{\tilde{\Upsilon}}) =-\mu_1(I_{q}\otimes \bar{\Psi}^D_{L} \otimes I_{p+q} ){\rm vec }(\tilde{\Upsilon}) $.
Consider the following Lyapunov function:
\begin{equation}
    V_1(t)={\rm vec}(\tilde{\Upsilon})^{\mathrm{T}} (\Theta \otimes \bar{P}_1) {\rm vec}(\tilde{\Upsilon}).
\end{equation}
The derivative of $V_1(t)$ can be calculated as
\begin{align}
    \dot{V}_1 =& {\rm vec}(\tilde{\Upsilon})^{\mathrm{T}}( -\mu_1 ((\bar{\Psi}^D_{L})^{\mathrm{T}} \Theta + \Theta \bar{\Psi}^D_{L} )\otimes \bar{P}_1) {\rm vec}(\tilde{\Upsilon}) \notag\\
    \leq&  -\mu_1 \lambda_{\rm min}(\Omega^D \Theta^{-1} ) V_1,
    \end{align}
where $ \Omega^D=(\bar{\Psi}^D_{L})^{\mathrm{T}} \Theta + \Theta \bar{\Psi}^D_{L} $.
Then, we obtain
%\begin{flalign}
\begin{equation}
    V_1(t)\leq {\rm e}^{-\mu_1 \lambda_{\rm min}(\Omega^D \Theta^{-1} )(t-t_0)} V_1(t_0),
\end{equation}
and
\begin{equation}
    {\rm vec} (\tilde{\Upsilon}) \leq \sqrt{ \frac{V_1(t_0)}{\lambda_{\rm min}(\Theta \otimes \bar{P}_1)} } {\rm e}^{-\frac{\mu_1}{2} \lambda_{\rm min}(\Omega^D \Theta^{-1} ) (t-t_0)}.
\end{equation}
%\end{flalign}
From Definition \ref{Def3}, we have
\begin{align}
  \bar{\Psi}_{L}^D(t-t_0) =& \bar{\Psi}_{L}\left\lvert \Xi_N(t_0,t)\right\rvert \notag\\
  =&\bar{\Psi}_{L}|t-t_0-\Xi_A(t_0,t)|\notag\\
  \geq & \bar{\Psi}_{L} \left(t-t_0-\left(T_0+ \frac{t-t_0}{\tau_a} \right) \right)\notag\\
  %\geq & \bar{\Psi}_{L}((1-\frac{1}{\tau_a})(t-t_0)+T_0) \\
  \geq & \left(1-\frac{1}{\tau_a}\right) \bar{\Psi}_{L} (t-t_0)  ,~ t\geq t_0.
\end{align}
Next, it is easy to get that $\lambda_{\rm min}(\Omega^D \Theta^{-1} ) (t-t_0) \geq \left(1-\frac{1}{\tau_a}\right) \lambda_{\rm min}(\Omega \Theta^{-1} ) (t-t_0).$

Therefore, we conclude that
\begin{equation}
    {\rm vec} (\tilde{\Upsilon}) \leq c_{\Upsilon} {\rm e}^{-\alpha_{\Psi}(t-t_0)},
\end{equation}
where $c_{\Upsilon}= \sqrt{ \frac{V_1(t_0)}{\lambda_{\rm min}(\Theta \otimes \bar{P}_1)}}$ and $\alpha_{\Psi}=\frac{ \mu_1}{2} (1-\frac{1}{\tau_a}) \lambda_{\rm min}(\Omega \Theta^{-1} ) $. It implies that ${\rm vec}(\tilde{\Upsilon})$ and $\tilde{\Upsilon}$ converge to zero exponentially at rates of at least $\alpha_{\Psi}$. Moreover, since $\alpha_{\Psi}>\sigma_{\rm max}(S)$, $\tilde{S}_i x_k$ converges to zero exponentially for $i\in \mathcal{F}, k\in \mathcal{L}$.
$\hfill \hfill \blacksquare $

\begin{remark}
In practice, various noises and attacks exist among agents. Therefore, we reconstruct the leader dynamics of all followers according to the observers (\ref{EQ15}). The observer gain $\mu_1$ is designed to guarantee that the convergence rate of the estimation value $\tilde{S}$ is greater than the divergence rate of the leader state. Moreover, $\mu_1$ guarantees the convergence of the containment error.
Compared with the previous work \cite{chen2019}, which only considered the observers in the consensus of a single leader under bounded actuator attacks, this paper considers heterogeneous output containment under composite attacks.
$\hfill \hfill \square $
\end{remark}
\subsection{Design of Distributed Resilient State Estimators}
Based on the reconstructed leader dynamics, distributed resilient estimators are proposed to estimate follower states under DoS attacks.
Consider the following distributed state estimators on the TL:
\begin{flalign}\label{equation 200}
 & \dot{\zeta}_i \!=\!\!\hat{S}_i \zeta_i \!+\!\mu_2 G \!\left(\sum _{j \in \mathcal{F} }d_{ij}(\zeta_j-\zeta_i)\!+\!\sum_{k \in \mathcal{L}}d_{ik}(\zeta_k-\zeta_i)\!\right)\!\!,&
\end{flalign}
%{\color{blue}
where $\zeta_k=x_k$, $\zeta_i$ is the $i$th follower state estimate on the TL, $\mu_2  > 0$ is the estimator gain, which can be chosen by the user, and $G$ is the  coupling gain to be determined later.
%}
The global state of the TL can be represented as
\begin{equation}
  \dot{\zeta}= \hat{S}_b \zeta-\mu_2 G_b \left(\sum_{k \in \mathcal{L}}(\Psi_k^D \otimes I_p )(\zeta-\bar{\zeta}_k)\right),
\end{equation}
where $\hat{S}_b={\rm blkdiag}(\hat{S}_i) $, $G_b=I_N \otimes G$, $\zeta=[\zeta_1^{\mathrm{T}},\zeta_2^{\mathrm{T}},\dots,\zeta_N^{\mathrm{T}}]^{\mathrm{T}}$, and $\bar{\zeta}_k=l_n \otimes \zeta_k$.

Define the global state estimation error of the TL as
\begin{equation}
\begin{aligned}
  \tilde{\zeta}\!
  =&\zeta-\left(\sum_{r \in \mathcal{L}}(\Psi_r \otimes I_p )\right)^{-1} \sum_{k \in \mathcal{L}}(\Psi_k\otimes I_p ) \bar{\zeta}_k.
\end{aligned}
\end{equation}
Then, during the normal communication, we have
\begin{equation}\label{EQ26}
\begin{aligned}
      \dot{\tilde{\zeta}}\!
       =&\hat{S}_b \zeta-\mu_2 (I_N \otimes G) \left(\sum_{k \in \mathcal{L}}(\Psi_k \otimes I_p )(\zeta-\bar{\zeta}_k)\right) \\
       &-
  \left(\sum_{r \in \mathcal{L}}(\Psi_r \otimes I_p )\right)^{-1}\sum_{k \in \mathcal{L}}(\Psi_k \otimes I_p ) (I_n \otimes S) \bar{\zeta}_k \\
      =&\hat{S}_b \zeta- (I_n \otimes S)  \zeta+ (I_n \otimes S)  \zeta \\
  &-(I_n \otimes S)\left(\sum_{r \in \mathcal{L}}(\Psi_r \otimes I_p )\right)^{-1} \sum_{k \in \mathcal{L}}(\Psi_k \otimes I_p )  \bar{\zeta}_k  +M \\
  & -\mu_2 (I_N \otimes G) \sum_{k \in \mathcal{L}}(\Psi_k \otimes I_p )\\
  & \Bigg( \zeta -\left(\sum_{\bar{r} \in \mathcal{L}}(\Psi_{\bar{r}} \otimes I_p )\right)^{-1} \sum_{\bar{k} \in \mathcal{L}}(\Psi_{\bar{k}} \otimes I_p )\bar{\zeta}_{\bar{k}}\\
  &+
  \left(\sum_{\bar{r} \in \mathcal{L}} (\Psi_{\bar{r}} \otimes I_p )\right)^{-1} \sum_{\bar{k} \in \mathcal{L}}(\Psi_{\bar{k}} \otimes I_p )\bar{\zeta}_{\bar{k}}-  \bar{\zeta}_k \Bigg)\\
  =&\tilde{S}_b \zeta+(I_n \otimes S) \tilde{\zeta}-\mu_2 (I_N \otimes G)\sum_{k \in \mathcal{L}}(\Psi_k \otimes I_p ) \tilde{\zeta} \!+\!M \\
  &-
  \mu_2 (I_N \otimes G) \! \left(\! \sum_{\bar{k} \in \mathcal{L}} \! (\Psi_{\bar{k}} \otimes I_p ) \bar{\zeta}_{\bar{k}}-\! \sum_{k \in \mathcal{L}} \! (\Psi_k \otimes I_p )\bar{\zeta}_k \! \right)  \\
  =&(I_n \otimes S) \tilde{\zeta} \!-\!\mu_2 (I_N \otimes G)\sum_{k \in \mathcal{L}}(\Psi_k \otimes I_p ) \tilde{\zeta}\!+\! \tilde{S}_b \tilde{\zeta} \!+\! F_{\zeta}(t),&
    \end{aligned}
\end{equation}
where  $F_{\zeta}(t)=\tilde{S}_b\left(\sum_{r \in \mathcal{L}}(\Psi_r \otimes I_p )\right)^{-1} \sum_{k \in \mathcal{L}}(\Psi_k \otimes I_p ) \bar{\zeta}_k +M$ and
$M = (I_n \otimes S) \left(\sum_{r \in \mathcal{L}}(\Psi_r \otimes I_p )\right)^{-1}\sum_{k \in \mathcal{L}}(\Psi_k \otimes I_p ) \bar{\zeta}_k - \left(\sum_{r \in \mathcal{L}}(\Psi_r \otimes I_p )\right)^{-1}\sum_{k \in \mathcal{L}}(\Psi_k \otimes I_p ) (I_n \otimes S) \bar{\zeta}_k$ and $\tilde{S}_b={\rm blkdiag}(\tilde{S}_i)$ for $i\in \mathcal{F}$.

During the denied communication, we have
\begin{flalign}\label{EQ27}
      \dot{\tilde{\zeta}}\!
       =&\hat{S}_b \zeta-
  \left(\sum_{r \in \mathcal{L}}(\Psi_r \otimes I_p )\right)^{-1}\sum_{k \in \mathcal{L}}(\Psi_k \otimes I_p ) (I_n \otimes S) \bar{\zeta}_k &\notag\\
      =&\hat{S}_b \zeta- (I_n \otimes S)  \zeta+ (I_n \otimes S)  \zeta &\notag\\
  &\!-\!(I_n \otimes S)\left(\sum_{r \in \mathcal{L}}(\Psi_r \otimes I_p )\right)^{-1} \sum_{k \in \mathcal{L}}(\Psi_k \otimes I_p )  \bar{\zeta}_k  +M &\notag\\
  %=&\tilde{S}_b \zeta+(I_n \otimes S) \tilde{\zeta} +M\\
  =&(I_n \otimes S) \tilde{\zeta}+ \tilde{S}_b \tilde{\zeta} + F_{\zeta}(t).&
    \end{flalign}
Therefore, according to (\ref{EQ26}) and (\ref{EQ27}), we can conclude that
\begin{equation}\label{EQ32}
    \dot{\tilde{\zeta}} = \begin{cases}
    \begin{aligned}
    \hat{S}_b \tilde{\zeta}-\mu_2 (I_N \otimes G)\sum_{k \in \mathcal{L}}
    (\Psi_k \otimes I_p )  \tilde{\zeta} + F_{\zeta}(t),\\
    ~t \in \Xi_N(t_0,t),
    \end{aligned} \\
     \hat{S}_b \tilde{\zeta} + F_{\zeta}(t),~ t \in \Xi_A(t_0,t).
    \end{cases}
\end{equation}

\begin{Lemma}\label{Lemma5}
Under Lemma 2 and the Kronecker product property $(P \otimes Q)(Y \otimes Z) =(PY)\otimes(QZ)$, the following relation holds:

$\left(\sum_{r \in \mathcal{L}}(\Psi_r \otimes I_p )\right)^{-1}\sum_{k \in \mathcal{L}}(\Psi_k \otimes I_p ) (I_n \otimes S) \bar{\zeta}_k=
(I_n \otimes S) \left(\sum_{r \in \mathcal{L}}(\Psi_r \otimes I_p )\right)^{-1}\sum_{k \in \mathcal{L}}(\Psi_k \otimes I_p ) \bar{\zeta}_k$.

\textbf{Proof:}
Let
\begin{flalign}
M= &\sum_{k \in \mathcal{L}} M_k
=\sum_{k \in \mathcal{L}} \Bigg((I_n \otimes S) \left(\sum_{r \in \mathcal{L}}(\Psi_r \otimes I_p )\right)^{-1}(\Psi_k \otimes I_p ) &\notag\\
&-\left(\sum_{r \in \mathcal{L}}(\Psi_r \otimes I_p )\right)^{-1} (\Psi_k \otimes I_p ) (I_n \otimes S)\Bigg) \bar{\zeta}_k.&
\end{flalign}
According to the Kronecker product property $(P \otimes Q)(Y \otimes Z) =(PY)\otimes(QZ)$, we have
\begin{align}
  & (I_N \otimes S)\left(\sum_{r \in \mathcal{L}} \Psi_r \otimes I_p\right)^{-1} (\Psi_k \otimes I_p) \notag\\
   =& (I_N \otimes S)\left(\left(\sum_{r \in \mathcal{L}} \Psi_r\right)^{-1} \Psi_k\right) \otimes I_p \notag\\
   =&\left(I_N \times \left(\sum_{r \in \mathcal{L}} \Psi_r\right)^{-1} \Psi_k\right)\otimes(S \times I_p)\notag \\
   =& \left(\sum_{r \in \mathcal{L}} \Psi_r \otimes I_p\right)^{-1} (\Psi_k \otimes I_p)  (I_N \otimes S).
\end{align}
Therefore, $M_k=0$, which yields
\begin{equation}
     M=\sum_{k \in \mathcal{L}}M_k =0.
 \end{equation}
This completes the proof.
$\hfill \hfill \blacksquare $
\end{Lemma}

%}

%Editor: Please ensure that the intended meaning has been maintained in the following edit.
Then, by Lemma \ref{Lemma5} and Theorem 1, we obtain that $F_{\zeta}(t)=\tilde{S}_b\left(\sum_{r \in \mathcal{L}}(\Psi_r \otimes I_p )\right)^{-1} \sum_{k \in \mathcal{L}}(\Psi_k \otimes I_p ) \bar{\zeta}_k$ exponentially converges to zero.

\begin{myTheo}\label{Theorem 2}
    Consider the MAS (\ref{EQ1})-(\ref{EQ2}) under DoS attacks. Suppose that Assumption 1 holds and Definitions 2 and 3 are satisfied under DoS attacks. Then, there exist scalars
    $\tilde{\alpha}_1>0$, $\tilde{\alpha}_2>0$, and $\mu_2>0$ and a positive definite symmetric matrix $\bar{P}_2 >0$ such that
    \begin{align}
    \bar{P}_2 S+S^{\mathrm{T}}\bar{P}_2 - \bar{P}_2^{\mathrm{T}} \mu_2^2 \bar{P}_2 + \tilde{\alpha}_1 \bar{P}_2 =0, \\
    \bar{P}_2 S+S^{\mathrm{T}}\bar{P}_2  - \tilde{\alpha}_2 \bar{P}_2 \leq 0, \\
    \frac{1}{\tau_a} < \frac{\alpha_1}{\alpha_1+\alpha_2}, \label{EQ36}
    \end{align}
%where  $\bar{\Psi}_{L}=\sum_{k \in \mathcal{L}}(\Psi_k ) $,
where $\alpha_1=\tilde{\alpha}_1-k_1 ||\Theta \otimes \bar{P}_2||$, $\alpha_2=\tilde{\alpha}_2+k_1 ||\Theta \otimes \bar{P}_2||$ with $k_1>0$, and $G=\mu_2 \lambda_{\rm max}(\Omega^{-1} \Theta) \bar{P}_2$.
Then, the global state estimation error $\tilde{\zeta}$ of TL converges to zero exponentially under DoS attacks.
\end{myTheo}

\textbf{Proof.}
We prove that the TL state can achieve containment under DoS attacks.
For clarity, we first redefine the set $\Xi_A[t_0, t)$ as $\Xi_A[t_0, t)= \bigcup_{k=0,1,2,\dots} [t_{2k+1},t_{2k+2})$, where $t_{2k+1}$ and $t_{2k+2}$ indicate the times when the attacks start and end, respectively.
Then, the set $\Xi_N[t_0,t)$ can be redefined as $\Xi_N[t_0, t)= \bigcup_{k=0,1,2,\dots} [t_{2k},t_{2k+1})$.

Consider the following Lyapunov function candidate:
\begin{align}
   V_2(t)=\tilde{\zeta}^{\mathrm{T}} (\Theta \otimes \bar{P}_2) \tilde{\zeta}.
\end{align}
During the normal communication, the time derivative of $V_2(t)$ is computed as
\begin{flalign}
\dot{V}_2
=&\tilde{\zeta}^{\mathrm{T}}(\Theta \otimes (\bar{P}_2 S + S^{\mathrm{T}}\bar{P}_2))\tilde{\zeta} &\notag\\
&- \tilde{\zeta}^{\mathrm{T}}
((\bar{\Psi}^{\mathrm{T}}_L \Theta  +\Theta  \bar{\Psi}_L) \otimes \lambda_{\rm max}(\Omega^{-1} \Theta) \mu_2^2 \bar{P}_2^2 )\tilde{\zeta} &\notag\\
&+\tilde{\zeta}^{\mathrm{T}} (\Theta \otimes (\bar{P}_2 \tilde{S}_i + \tilde{S}_i^{\mathrm{T}}\bar{P}_2) ) \tilde{\zeta}  +2\tilde{\zeta}^{\mathrm{T}} (\Theta \otimes \bar{P}_2) F_{\zeta} &\notag\\
\leq& \tilde{\zeta}^{\mathrm{T}} ((\Theta \otimes \bar{P}_2) \tilde{S}_b+ \tilde{S}_b^{\mathrm{T}} (\Theta \otimes \bar{P}_2)) \tilde{\zeta} +2\tilde{\zeta}^{\mathrm{T}} (\Theta \otimes \bar{P}_2) F_{\zeta}&\notag\\
&+\tilde{\zeta}^{\mathrm{T}} (\Theta \otimes (\bar{P}_2 S + S^{\mathrm{T}}\bar{P}_2) - \Theta \otimes \mu_2^2 \bar{P}_2^2)\tilde{\zeta}.&
\end{flalign}
By Young's inequality, there exists a scalar $k_1>0$ that yields
\begin{align} \label{EQ39}
    2\tilde{\zeta}^{\mathrm{T}}(\Theta \otimes \bar{P}_2) F_{\zeta} \leq k_1 \tilde{\zeta}^{\mathrm{T}} (\Theta \otimes \bar{P}_2)^2 \tilde{\zeta} + \frac{1}{k_1} ||F_{\zeta}||^2 .
\end{align}
Then, we have
\begin{flalign}\label{EQ 41}
\dot{V}_2\leq& -\tilde{\alpha}_1 V_2 + 2  ||\tilde{S}_b|| V_2 +k_1 ||\Theta \otimes \bar{P}_2||V_2 +\frac{1}{k_1} ||F_{\zeta}||^2&\notag\\
=&\left(-\tilde{\alpha}_1+k_1 ||\Theta \otimes \bar{P}_2||+2  ||\tilde{S}_b||\right)V_2+\frac{1}{k_1} ||F_{\zeta}||^2.&
\end{flalign}
Because $\tilde{S}_b$ and $F_{\zeta}$ converge to zero exponentially, we get
\begin{equation} \label{EQ45}
    \dot{V}_2\leq  (-\alpha_1+ a_s {\rm e}^{-b_s t})V_2 + \varphi(t),
\end{equation}
where $a_s>0$, $b_s>0$, and $\varphi(t)=a_f {\rm e}^{-b_f t}$ with $a_f>0$ and $b_f>0$.
When communication is denied, the time derivative of $V_2(t)$ is calculated as
\begin{flalign}\label{EQ42}
\dot{V}_2
=&\tilde{\zeta}^{\mathrm{T}}(\Theta \otimes (\bar{P}_2 S + S^{\mathrm{T}}\bar{P}_2))\tilde{\zeta}
+ \tilde{\zeta}^{\mathrm{T}}
((\Theta \otimes \bar{P}_2) \tilde{S}_b &\notag\\
&+ \tilde{S}_b^{\mathrm{T}} (\Theta \otimes \bar{P}_2)) \tilde{\zeta} +2\tilde{\zeta}^{\mathrm{T}} (\Theta \otimes \bar{P}_2) F_{\zeta} &\notag\\
\leq& \tilde{\alpha}_2 V_2 + 2||\tilde{S}_b||  V_2 +k_1 ||\Theta \otimes \bar{P}_2||V_2 +\frac{1}{k_1} ||F_{\zeta}||^2 &\notag\\
 \leq& (\alpha_2+ a_s {\rm e}^{-b_s t}) V_2 + \varphi(t). &
\end{flalign}
By solving inequalities (\ref{EQ45}) and (\ref{EQ42}), we obtain the following inequality:
\begin{equation} \label{EQ47}
   V_2(t)  \! \leq  \! \begin{cases}\begin{aligned}
   & {\rm e}^{\int_{t_{2k}}^{t}-\alpha_1+ a_s {\rm e}^{-b_s \tau}\,{\rm d}\tau}V_2(t_{2k}) \\
   & \!+ \!\int_{t_{2k}}^{t} {\rm e}^{\int_{\tau}^{t}-\alpha_1+ a_s {\rm e}^{-b_s s}\,{\rm d}s} \varphi(\tau)\,{\rm d}\tau ,~ t \! \in  \! [t_{2k},t_{2k+1}),
   \end{aligned}\\
   \begin{aligned}
    & {\rm e}^{\int_{t_{2k+1}}^{t}\alpha_2+ a_s {\rm e}^{-b_s \tau}\,{\rm d}\tau} V_2(t_{2k+1}) \\
    &\!+ \!\int_{t_{2k+1}}^{t}\! \! {\rm e}^{\int_{\tau}^{t}\!\alpha_2+ a_s {\rm e}^{-b_s s}\!\,{\rm d}s} \varphi(\tau)\!\,{\rm d}\tau ,~ t \! \in  \! [t_{2k+1},t_{2k+2}) .
   \end{aligned}
   \end{cases}
\end{equation}
Denoting $\alpha=\begin{cases}
 -\alpha_1,~ t\in[t_{2k},t_{2k+1}),\\
 \alpha_2,~t \in[t_{2k+1},t_{2k+2}),
\end{cases}$  we rewrite (\ref{EQ47}) as
\begin{flalign}\label{EQ46}
    &V_2(t)\! \leq \! {\rm e}^{\int_{t_0}^{t}\!\alpha  + a_s {\rm e}^{-b_s \tau}\! \,{\rm d}\tau}V_2(t_{0}) \!+\! \int_{t_{0}}^{t} \!{\rm e}^{\int_{\tau}^{t}\!\alpha + a_s {\rm e}^{-b_s s}\! \,{\rm d}s}\! \varphi(\tau)\,{\rm d}\tau .&
\end{flalign}
For $t\in \Xi_{N}[t_0,t)$, (\ref{EQ46}) implies that
\begin{equation}\label{EQ548}
\begin{aligned}
   V_2(t)\! \leq& {\rm e}^{\int_{t_0}^{t} a_s {\rm e}^{-b_s \tau}\, d\tau} {\rm e}^{\int_{t_0}^{t}\alpha \,{\rm d}\tau}V_2(t_{0}) \\
   &+ \int_{t_{0}}^{t}          a_f {\rm e}^{-b_f+ \int_{\tau}^{t} a_s {\rm e}^{-b_s s}\,{\rm d} s }  {\rm e}^{\int_{\tau}^{t}\alpha\,{\rm d}s} \,{\rm d}\tau \\
    \leq& {\rm e}^{-\frac{a_s}{b_s}({\rm e}^{-b_s t} -{\rm e}^{-b_s t_0})}  {\rm e}^{-\alpha_1|\Xi_N(t_0,t)|+\alpha_2|\Xi_A(t_0,t)|}V_2(t_0)\\
    &\!+\!\int_{t_0}^{t}\! a_f \! {\rm e}^ { \!-\! b_f \tau\!-\!\frac{a_s}{b_s}({\rm e}^{\!-\!b_s t}\! -\!{\rm e}^{-b_s \tau}\!)} {\rm e}^{\!-\!\alpha_1|\Xi_N(\tau,t)|\!+\!\alpha_2|\Xi_A(\tau,t)|}\,{\rm d} \tau .
\end{aligned}
\end{equation}

Similarly, it can be concluded that (\ref{EQ548}) holds for $t\in \Xi_{A}[t_0,t)$. According to Definition \ref{Def3}, we have
\begin{flalign}\label{EQ49}
   & -\alpha_1|\Xi_N(t_0,t)+\alpha_2|\Xi_A(t_0,t)| &\notag\\
    =& -\alpha_1 \left(t-t_0-|\Xi_A(t_0,t)|\right) +\alpha_2|\Xi_A(t_0,t)| &\notag\\
    \leq& -\alpha_1 (t-t_0)+(\alpha_1+\alpha_2)  \left(T_0+\frac{t-t_0}{\tau_a} \right) &\notag\\
    \leq& -\eta(t-t_0)+(\alpha_1+\alpha_2)T_0,&
\end{flalign}
where $\eta =(\alpha_1  - \frac{\alpha_1+\alpha_2}{\tau} )$.
Substituting (\ref{EQ49}) into (\ref{EQ548}), we obtain
\begin{flalign}
   V_2(t)
    \leq& {\rm e}^{-\frac{a_s}{b_s}({\rm e}^{-b_s t}-{\rm e}^{-b_s t_0})+(\alpha_1+\alpha_2)T_0}V_2(t_0){\rm e}^{-\eta(t-t_0)} &\notag\\
    &+\frac{a_f}{-b_f+\eta}  {\rm e}^{-\frac{a_s}{b_s}({\rm e}^{-b_s t}-{\rm e}^{-b_s t_0})+(\alpha_1+\alpha_2)T_0} &\notag\\
    &({\rm e}^{-b_f t}-{\rm e}^{-\eta t+(-b_f +\eta)t_0}) &\notag\\
   \leq& c_1 {\rm e}^{-\eta(t-t_0)}+c_2 {\rm e}^{-b_f t}&
    \end{flalign}
with $c_1={\rm e}^{\frac{a_s}{b_s}{\rm e}^{-b_s t_0}+(\alpha_1+\alpha_2)T_0}\left(V_2(t_0)-\frac{a_f}{-b_f+\eta}{\rm e}^{-b_f t_0}\right)$ and $c_2= \frac{a_f}{-b_f+\eta} {\rm e}^{\frac{a_s}{b_s}{\rm e}^{-b_s t_0}+(\alpha_1+\alpha_2)T_0}$.
From (\ref{EQ36}), we obtain $\eta>0$.
{Therefore, $V_2(t)$ and $\tilde{\zeta}$ converge to zero exponentially. }

%From (\ref{EQ36}), we obtain $\eta>0$. Therefore, $V_2(t)$ and $\tilde{\zeta}$ converge to zero exponentially. Thus, we have proved that $\tilde{\zeta}$ converges to zero,

%\footnote{Thus, we prove that $\tilde{\zeta}$ converges to zero, which is bounded. {\color{blue}Moreover, the inequality (\ref{EQ39}) can be rewritten as $ 2\tilde{\zeta}^{\mathrm{T}}(\Theta \otimes \bar{P}_2) F_{\zeta} \leq a_{f}^* {\rm e}^{-b_{f}^* t}$ with $a_{f}^*>0$ and $b_{f}^*>0$. Then, similarly to (\ref{EQ 41})-(\ref{EQ42}), it can calculate that $\dot{V}_2  \leq \left(-\tilde{\alpha}_1+2  ||\tilde{S}_b||\right)V_2+a_{f}^* {\rm e}^{-b_{f}^* t} \leq (-\tilde{\alpha}_1+ a_s {\rm e}^{-b_s t})V_2 + \varphi^{\ast}(t)$ with $\varphi^{\ast}(t)=a_{f}^* {\rm e}^{-b_{f}^* t} $ for the normal communication and $\dot{V}_2 \leq (\tilde{\alpha}_2+ a_s {\rm e}^{-b_s t})V_2 + \varphi^{\ast}(t)$ for denied communication. Then, for (43)-(46), $\alpha_1$ and $\alpha_2$ can be written as $\tilde{\alpha}_1$ and $\tilde{\alpha}_2$, which can get that $\eta^{\ast} =(\tilde{\alpha}_1  - \frac{\tilde{\alpha}_1+\tilde{\alpha}_2}{\tau} ) >0$, then, (\ref{EQ36}) can be rewritten as $\frac{1}{\tau_a} < \frac{\tilde{\alpha}_1}{\tilde{\alpha}_1+\tilde{\alpha}_2}$, which shows that  $\alpha_1=\tilde{\alpha}_1$ and $\alpha_2=\tilde{\alpha}_2$ hold.}}

\begin{remark}
Compared with previous results \cite{deng2021}, our state estimators can work without knowing accurate leader dynamics, which is more appropriate in practical applications.
In \cite{deng2021}, Deng et al. considered a single leader consensus, whereas we address a more general case of the output containment with multiple leaders. In addition, in contrast to \cite{yang2020}, where the state tracking errors are only bounded under DoS attacks, our papers prove that the TL state error converges to zero exponentially.
$\hfill \hfill \square $
\end{remark}

%}

\subsection{Decentralized Output Regulator Equation Solvers}
Our control protocol uses the output regulator equations to provide an appropriate feedforward control gain and achieve the output containment. However, the output regulator equations require knowledge of  the leader dynamics. Since MASs are fragile, various disturbances and attacks can occur in the information transmission channel of the CPL.
Therefore, we cannot use the leader dynamics matrices $S$ and $R$ directly, as discussed in \cite{zuo2020}. Since the leader dynamics are reconstructed on the TL, we use the reconstructed dynamic matrices $\hat{S}_i$ and $\hat{R}_i$ to solve the output regulator equations according to the following theorem.

\begin{myTheo}
Suppose that Assumptions \ref{assumption 1} and \ref{assumption 4} hold. The estimated solutions $\hat{\Delta}_{i}$  to the output regulator equations in (\ref{EQ10}) can be obtained as
\begin{align} \label{EQ48}
    &\dot{\hat{\Delta}}_{i} = - \mu_3 \hat{\Phi }^{\mathrm{T}}_i(\hat{\Phi }_i \hat{\Delta}_{i}-\hat{\mathcal{R}}_i),
\end{align}
%\begin{equation}
%   \dot{\hat{\Delta}}_{i} = - u \hat{\Phi }^{\mathrm{T}}_i(\hat{\Phi }_i \hat{\Delta}_{i}-\hat{\mathcal{R}}_i)
%\end{equation}
%with the gain $\mu_3 > \frac{\sigma_{\rm max}(S)}{\lambda_{\rm min}(\Phi_i^{\mathrm{T}} \Phi_i )}$, then, the estimated solution $\hat{\Delta}_i$ exponentially converge to $\Delta$ .
where $\hat{\Delta}_i={\rm vec}(\hat{Y}_i)$, $\hat{\Phi}_i=(I_q \otimes  M_i-\hat{S}_i^{\mathrm{T}} \otimes N_i)$, $\mathcal{\hat{R}}_i={\rm vec}(\hat{R}^{\ast}_i)$,  $\hat{Y_{i}}=\left[\begin{array}{cc}
     \hat{\Pi}_{i}  \\
     \hat{\Gamma}_{i}
\end{array}\right]$,  $\hat{\Gamma}_i$ and $\hat{\Pi}_i$  are estimates of $\Gamma$ and $\Pi$ in (\ref{EQ10}), $\hat{R}^{\ast}_i=\left[\begin{array}{cc}
     0  \\
     \hat{R}_i
\end{array}\right]$, $
M_i=\left[
  \begin{array}{cc}
  A_i &  B_i  \\
  C_i &  0   \\
  \end{array}
  \right]$, $N_i=
    \left[
  \begin{array}{cc}
   I_{n_i} & 0  \\
   0 &  0   \\
  \end{array}
  \right]$, and $\mu_3 > \frac{\sigma_{\rm max}(S)}{\lambda_{\rm min}(\Phi_i^{\mathrm{T}} \Phi_i )}$ with $\Phi_i=(I_q \otimes  M_i-S^{\mathrm{T}} \otimes N_i)$.
\end{myTheo}

\textbf{Proof.}
Using Assumption \ref{assumption 4}, we obtain that the output regulator equations:
 \begin{equation}\label{EQ10}
     \begin{cases}
      A_i\Pi_i+B_i\Gamma_i=\Pi_i S, \\
      C_i\Pi_i = R,
     \end{cases}
 \end{equation}
have a pair of unique solution matrices $\Pi_i$ and $\Gamma_i$ for each $i=\textbf{I}[1, N]$.

Rewriting the output regulator equation (\ref{EQ10}) yields
\begin{flalign}
 &\left[
  \begin{array}{cc}
  A_i &  B_i  \!\\
  C_i &  0 \!  \\
  \end{array}
  \right]\!\!
   \left[
    \begin{array}{cc}
   \Pi_i \\
   \Gamma_i \\
    \end{array}
    \right]\!
    I_q -\!
    \left[
  \begin{array}{cc}
   I_{n_i} & 0  \\
   0 &  0   \\
  \end{array}
  \right]\!\!
   \left[
    \begin{array}{cc}
\Pi_i\\
\Gamma_i \\
    \end{array}
    \right]\!S
    =\!\!
    \left[\begin{array}{cc}
        \! 0\!  \\
       \!  R\!
    \end{array}\right],&
\end{flalign}
which can be rewritten as
\begin{equation}\label{EQ511}
    M_iY I_q -N_iY S=R_i^{\ast},
\end{equation}
where $M_i=\left[
  \begin{array}{cc}
  A_i &  B_i  \\
  C_i &  0   \\
  \end{array}
  \right],
  N_i=\left[
  \begin{array}{cc}
   I_{n_i} & 0  \\
   0 &  0   \\
  \end{array}
  \right],
  Y_i= \left[
    \begin{array}{cc}
\Pi_i\\
\Gamma_i \\
    \end{array}
    \right], $ and $R_i^{\ast}= \left[\begin{array}{cc}
         0  \\
         R
    \end{array}\right] $. According to Theorem 1.9 of Huang et al. \cite{huang2004}, the linear equation (\ref{EQ511}) can be represented as
\begin{equation}\label{EQ52}
    \Phi_i \Delta_i=\mathcal{R}_i,
\end{equation}
where $\Phi_i=(I_q \otimes  M_i-S^{\mathrm{T}} \otimes N_i)$, $\Delta_i={\rm vec}( Y_i)$,
    $\mathcal{R}_i={\rm vec}(R^{\ast})$.

Similarly, using reconstructed leader dynamics, we obtain the output regulator equations:
 \begin{equation} \label{EQ 53}
     \begin{cases}
      A_i\hat{\Pi}_i+B_i\hat{\Gamma}_i=\hat{\Pi}_i \hat{S}_i, \\
      C_i\hat{\Pi}_i = \hat{R}_i,
     \end{cases}
 \end{equation}
and
\begin{equation}\label{EQ54}
   \hat{\Phi}_i \hat{\Delta}_i=\hat{\mathcal{R}}_i.
\end{equation}

In Theorem 1, we note that the estimated leader dynamics are time-varying, therefore, the output regulator equations influenced by the estimated leader dynamics are also time-varying. Next, we prove that the estimated solutions of the output regulator equations $\hat{\Delta}_{i}$ converge exponentially to the solutions of the output regulator equations $\Delta_i$.

Note that
\begin{flalign}
\dot{\hat{\Delta}}_{i}
=& - \mu_3 \hat{\Phi }^{\mathrm{T}}_i(\hat{\Phi }_i \hat{\Delta}_{i}-\hat{\mathcal{R}}_i) &\notag\\
=&- \mu_3  \hat{\Phi}^{\mathrm{T}}_i \hat{\Phi}_i \hat{\Delta}_{i}+  \mu_3 \hat{\Phi}_i \hat{\mathcal{R}_i} &\notag\\
=&-\!  \mu_3\Phi_i^{\mathrm{T}}\! \Phi_i \hat{\Delta}_{i} \!+\! \mu_3 \Phi_i^{\mathrm{T}} \!\Phi_i \hat{\Delta}_{i} \!-\! \mu_3 \hat{\Phi}_i^{\mathrm{T}}  \hat{\Phi}_i \hat{\Delta}_{i}\! +\!  \mu_3 \hat{\Phi}_i^{\mathrm{T}} \hat{\mathcal{R}_i} &\notag\\
&- \mu_3 \Phi_i^{\mathrm{T}} \hat{\mathcal{R}_i} +  \mu_3 \Phi_i^{\mathrm{T}} \hat{\mathcal{R}_i} - \mu_3 \Phi_i^{\mathrm{T}} \mathcal{R}_i +  \mu_3 \Phi_i^{\mathrm{T}} \mathcal{R}_i &\notag\\
=&-  \mu_3\Phi_i^{\mathrm{T}} \Phi_i \hat{\Delta}_{i} +  \mu_3 (\Phi_i^{\mathrm{T}} \Phi_i - \hat{\Phi}_i^{\mathrm{T}} \hat{\Phi}_i) \hat{\Delta}_{i} &\notag\\
&+   \mu_3 (\hat{\Phi}_i^{\mathrm{T}} - \Phi_i^{\mathrm{T}}) \hat{\mathcal{R}_i}
+  \mu_3 \Phi_i^{\mathrm{T}} (\hat{\mathcal{R}_i}-\mathcal{R}_i) +  \mu_3 \Phi_i^{\mathrm{T}} \mathcal{R}_i &\notag\\
=&- \mu_3 \Phi_i^{\mathrm{T}} \Phi_i \hat{\Delta}_{i} + \mu_3  \Phi_i^{\mathrm{T}} \mathcal{R}_i +d_i(t),&
\end{flalign}
where  $d_i(t)= - \mu_3 (\hat{\Phi}_i^{\mathrm{T}}\hat{\Phi}_i - \Phi_i^{\mathrm{T}}\Phi_i) \hat{\Delta}_{i} + \mu_3  \tilde{\Phi}_i^{\mathrm{T}} \hat{\mathcal{R}}_i +         \mu_3 \Phi_i^{\mathrm{T}} \tilde{\mathcal{R}}_i$ with $\tilde{\Phi}_i=\hat{\Phi}_i-\Phi_i=\tilde{S}_i^{\mathrm{T}} \otimes N_i$ and $\tilde{\mathcal{R}}={\rm vec}(\left[\begin{array}{cc}
    0  \\
    \tilde{R}_i
    \end{array}
    \right])$.

It is evident that $\lim _{t \rightarrow \infty}d_i(t) =0 $ exponentially at a rate of $\alpha_{\Psi}$.

Let $\tilde{\Delta}_{i}= \hat{\Delta}_{i}-\Delta_i $. The time derivative of $\tilde{\Delta}_{i}$
is calculated as
\begin{equation}\label{EQ53}
\begin{aligned}
    \dot{\tilde{\Delta}}_{i}
    =&- \mu_3 \Phi_i^{\mathrm{T}} \Phi_i \tilde{\Delta}_{i} -  \mu_3 \Phi_i^{\mathrm{T}} \Phi_i \Delta_i+  \mu_3\Phi_i^{\mathrm{T}} \mathcal{R}_i +d_i(t) \\
   =&- \mu_3 \Phi_i^{\mathrm{T}} \Phi_i \tilde{\Delta}_{i} +d_i(t).
\end{aligned}
\end{equation}
Solving (\ref{EQ53}), we obtain
\begin{equation}
\begin{aligned}
 \tilde{\Delta}_i(t)=\tilde{\Delta}_i(t_0){\rm e}^{-\mu_3 \Phi_i^{\mathrm{T}} \Phi_i(t-t_0)}+\int_{t_0}^{t} d_i(\tau) {\rm e}^{- \mu_3 \Phi_i^{\mathrm{T}} \Phi_i(t -\tau)}\,{\rm d}\tau .
\end{aligned}
\end{equation}

Since $ \Phi_i^{\mathrm{T}} \Phi_i$ is positive and $d_i(t)$ converges to zero at a rate of $\alpha_{\Psi}$, $\lim _{t \rightarrow \infty }\tilde{\Delta}_{i}=0$ exponentially. Moreover, because $\mu_3 > \frac{\sigma_{\rm max}(S)}{\lambda_{\rm min}(\Phi_i^{\mathrm{T}} \Phi_i )}$, the exponential convergence rate of $\lim _{t \rightarrow \infty }\tilde{\Delta}_{i}=0$ is larger than $\sigma_{\rm max}(S)$.
$\hfill \hfill \blacksquare $
\begin{Lemma}\label{Lemma 6}%[\cite{chen2019}]
The distributed leader dynamics of the observers (\ref{EQ15}) ensure that $\dot{\hat{\Pi}}_i$ and $\dot{\hat{\Pi}}_i  \zeta_k$ converge to zero exponentially.
\end{Lemma}

\begin{remark}
%Editor: Please ensure that the intended meaning has been maintained in the following edit.
Since $\dot{\hat{\Pi}}_i$ is a component of $\dot{\hat{\Delta}}_i$, the convergence rate of $\dot{\hat{\Pi}}_i$ is at least as large as that of $\dot{\tilde{\Delta}}_i$. Thus, the exponential convergence rate of $\dot{\hat{\Pi}}_i $ is larger than $\sigma_{\rm max}(S)$ according to $\mu_3 > \frac{\sigma_{\rm max}(S)}{\lambda_{\rm min}(\Phi_i^{\mathrm{T}} \Phi_i )}$. Then, it is clear that $\dot{\hat{\Pi}}_i  \zeta_k$ converges exponentially to zero.
%Compare with the existing work \cite{cai2017} that the gain $\mu_3 > \frac{\alpha_{\Psi}}{\lambda_{\rm min}(\Phi_i^{\mathrm{T}} \Phi_i )}$, our work have more conservative result.
$\hfill \hfill  \square $
\end{remark}
\subsection{Adaptive Decentralized Resilient Controllers Design}
Using the above theorems, we design a control scheme to achieve the output  containment synchronization of heterogeneous MASs under DoS attacks. Next, we combine the distributed observers, estimators and decentralized solvers, and adaptive control to achieve the containment resilience of heterogeneous MASs against DoS attacks and unknown unbounded actuation attacks (\ref{EQ9}).

Define the state tracking errors as
 \begin{align}
      \epsilon_i=x_i - \hat{\Pi}_i \zeta_i. \label{EQ58}
 \end{align}
Then, we design the decentralized adaptive attack-resilient control protocols on the CPL as follows:
\begin{align}
 & u_i=\hat{\Gamma}_i \zeta_i +K_i \epsilon_i -\hat{\chi}_i , \\
&\hat{\chi}_i=\frac{B_i^{\mathrm{T}}  P_i \epsilon_i}{\left\lVert \epsilon_i^{\mathrm{T}} P_i B_i \right\rVert +\omega} \hat{\rho_i}, \\ \label{EQ65}
&\dot{\hat{\rho}}_i=\begin{cases}
 \left\lVert \epsilon_i^{\mathrm{T}} P_i B_i \right\rVert +2\omega,& \mbox{if}  \left\lVert \epsilon_i^{\mathrm{T}} P_i B_i \right\rVert \geq \bar{d}, \\
 \left\lVert \epsilon_i^{\mathrm{T}} P_i B_i \right\rVert +2\omega \frac{\left\lVert \epsilon_i^{\mathrm{T}} P_i B_i \right\rVert}{\bar{d}},& \mbox{ otherwise},
\end{cases}
\end{align}
where $\hat{\chi}_i$ is an adaptive compensation signal, $\hat{\rho}_i$ is an adaptive updating parameter, $\omega$ is a given positive scalar, and the controller gain $K_i$ is assigned as
\begin{equation}\label{EQ0968}
    K_i=-R_i^{-1}B_i^{\mathrm{T}} P_i,
\end{equation}
where $P_i$ is the solution of
\begin{equation}\label{EQ 64}
    A_i^{\mathrm{T}} P_i + P_i A_i + Q_i -P_i B_i R_i^{-1} B_i^{\mathrm{T}} P_i =0.
\end{equation}

\begin{myTheo}
Consider a heterogeneous MAS (\ref{EQ1})-(\ref{EQ2}) with $M$ leaders and $N$ followers in presence of the composite attacks, including unbounded actuation attacks. Under Assumptions \ref{assumption 1}-\ref{assumption 5}, \textbf{Problem ACMCA} can be solved by designing the leader dynamic observers (\ref{EQ15}), distributed state estimators (\ref{equation 200}), decentralized output regulator equation solvers (\ref{EQ48}), and
%Editor: Please ensure that the intended meaning has been maintained in the following edit.
decentralized adaptive controllers (\ref{EQ58})-(\ref{EQ 64}).
\end{myTheo}

\textbf{Proof.}
By Theorem 1 and 2, we show that the TL can resist DoS attacks when attack frequency satisfies the conditions in Theorem 2. In the following, we show that the state tracking errors (\ref{EQ58}) are UUB under DoS attacks and unbounded actuation attacks. The derivative of $\epsilon_i$ is calculated as
  %\begin{equation}\label{EQ63}
  %\begin{aligned}
  \begin{flalign}\label{EQ63}
   \dot{\epsilon}_i
      =& A_i x_i + B_i u_i +B_i \chi_i -\dot{\hat{\Pi}}_i  \zeta_i &\notag \\
      &\! - \! \hat{\Pi}_i \!  \left( \! \hat{S}_i \zeta_i \!+\! \mu_2 G \!\left( \! \sum _{j \in \mathcal{F} } \!d_{ij}(\zeta_j-\zeta_i) \!+ \!\sum_{k \in \mathcal{L}}  \! d_{ik}(\zeta_k-\zeta_i  \!)\right)\!  \right) \! &\notag \\
      =&(A_i+B_i K_i)\epsilon_i  - \dot{\hat{\Pi}}_i \zeta_i &\notag\\
      &-\mu_2 \hat{\Pi}_i G  \! \left( \!  \sum _{j \in \mathcal{F} } \! d_{ij}(\zeta_j-\zeta_i) \! + \! \sum_{k \in \mathcal{L}} \! d_{ik}(\zeta_k-\zeta_i) \! \right) \! + \! B_i \tilde{\chi}_i,
  \end{flalign}
  %\end{aligned}
  %\end{equation}
where $\tilde{\chi}_i=\chi_i-\hat{\chi}_i$. The global state tracking error equation (\ref{EQ63}) can be presented as
 \begin{flalign}
 \dot{\epsilon}
  =&\bar{A}_b\epsilon -\dot{\hat{\Pi}}_b \zeta +\! \mu_2 \hat{\Pi}_b G_b \! \left( \! \sum_{k \in \mathcal{L}}\!(\Psi_k^D \!\otimes\! I_p )(\zeta \!-\!\bar{\zeta}_k) \! \right)\!+\!B_b\tilde{\chi}
 &\notag\\
 % =&\bar{A}_b\epsilon -\dot{\hat{\Pi}}_b(\tilde{\zeta}+ \left(\sum_{r \in \mathcal{L}}(\Psi_r \otimes I_p )\right)^{-1} \sum_{k \in \mathcal{L}}(\Psi_k \otimes I_p ) \bar{\zeta}_k) &\notag\\
 %& + \mu_2  \hat{\Pi}_b G_b \sum_{k \in \mathcal{L}}(\Psi_k^D \otimes I_p )\tilde{\zeta}
  %  +      \mu_2  \hat{\Pi}_b G_b \sum_{k \in \mathcal{L}}(\Psi_k^D \otimes I_p ) &\notag\\
  %  &(\left(\sum_{r \in \mathcal{L}}(\Psi_r \otimes I_p )\right)^{-1} \sum_{\bar{k} \in \mathcal{L}}(\Psi_{\bar{k}} \otimes I_p ) \bar\zeta_{\bar{k}}) -  \bar{\zeta}_k))     +        B_b \tilde{\chi}   &\notag \\
   =&\bar{A}_b\epsilon -\dot{\hat{\Pi}}_b \left( \tilde{\zeta}+ \left(\sum_{r \in \mathcal{L}}(\Psi_r \otimes I_p )\right)^{-1} \sum_{k \in \mathcal{L}}(\Psi_k \otimes I_p ) \bar{\zeta}_k \right) &\notag\\
   &+ \mu_2 \hat{\Pi}_b G_b \sum_{k \in \mathcal{L}}(\Psi_k^D \otimes I_p )\tilde{\zeta} +B_b \tilde{\chi},
 \end{flalign}
where $\bar{A}_b={\rm blkdiag}(A_i+B_i K_i)$, $B_b={\rm blkdiag}(B_i)$,  $\dot{\hat{\Pi}}_b={\rm blkdiag}(\dot{\hat{\Pi}}_i)$, $\hat{\Pi}_b= {\rm blkdiag}(\hat{\Pi}_i)$  for $i=\textbf{I}[1,N]$, and $\epsilon=[\epsilon_1^{\mathrm{T}}, \epsilon_2^{\mathrm{T}}, \dots ,\epsilon_N^{\mathrm{T}}]^{\mathrm{T}}$, $\tilde{\chi}=[\tilde{\chi}_1^{\mathrm{T}}, \tilde{\chi}_2^{\mathrm{T}}, \dots, \tilde{\chi}_N^{\mathrm{T}}]^{\mathrm{T}}$. Consider the following Lyapunov function:
\begin{equation} \label{EQ64}
    V= \epsilon ^{\mathrm{T}} P_b \epsilon,
\end{equation}
where $P_b={\rm blkdiag}(P_i)$. The time derivative of (\ref{EQ64}) is calculated as
\begin{flalign}\label{EQ80}
    \dot{V}
    =& 2\epsilon^{\mathrm{T}} P_b \dot{\epsilon}&\notag \\
    =&-\epsilon^{\mathrm{T}} Q_b \epsilon +2\epsilon^{\mathrm{T}} P_b \left(\dot{\hat{\Pi}}_b+  \mu_2 \hat{\Pi}_b G_b \sum_{k \in \mathcal{L}}(\Psi_k^D \otimes I_p ) \right)\tilde{\zeta}&\notag \\
    & \!+ \!2\epsilon^{\mathrm{T}}  P_b \dot{\hat{\Pi}}_b  \! \left(  \sum_{r \in \mathcal{L}}(\Psi_r \otimes I_p ) \right) \! ^ {-1} \! \sum_{k \in \mathcal{L}} (\Psi_k \otimes I_p ) \bar{\zeta}_k\! &\notag\\
    & +\! 2\epsilon^{\mathrm{T}} \!P_b B_b\tilde{\chi}&\notag\\
    \leq& -\sigma_{\rm min}(Q_b) \left\lVert \epsilon\right\rVert^2
    + 2\epsilon^{\mathrm{T}} P_b B_b\tilde{\chi}&\notag\\
    &+2\left\lVert \epsilon^{\mathrm{T}}\right\rVert \left\lVert P_b\right\rVert  \left\lVert \dot{\hat{\Pi}}_b+  \mu_2 \hat{\Pi}_b G_b \sum_{k \in \mathcal{L}}(\Psi_k \otimes I_p )  \right\rVert  \left\lVert \tilde{\zeta}\right\rVert&\notag\\
 & \! + \! 2  \! \left\lVert  \epsilon^{\mathrm{T}}\right\rVert  \! \left\lVert \!  P_b\right\rVert \!   \left\lVert \!  \dot{\hat{\Pi}}_b  \!
  \left(\sum_{r \in \mathcal{L}}(\Psi_r \otimes I_p ) \! \right) \! ^{-1} \! \sum_{k \in \mathcal{L}} \! (\Psi_k \otimes I_p )   \bar{\zeta}_k \! \right\lVert,
\end{flalign}
where $Q_b={\rm blkdiag}(Q_i)$. Since $\dot{\hat{\Pi}}_i $ and $ \dot{\hat{\Pi}}_i \zeta_k $ converge to zero exponentially in view of Lemma \ref{Lemma 6}, we have
\begin{equation}\label{EQ81}
    \left\lVert  \dot{\hat{\Pi}}_b
  \left(\sum_{r \in \mathcal{L}}(\Psi_r \otimes I_p )\right)^{-1}\sum_{k \in \mathcal{L}}(\Psi_k \otimes I_p ) \bar{\zeta}_k\right\rVert
  \leq  V_{\Pi} \exp (-\alpha_{\Pi})
\end{equation}
with $ V_{\Pi}>0$ and $\alpha_{\Pi}>0$.
Using Young's inequality, we obtain
\begin{equation}\label{EQ61}
\begin{aligned}
    & 2\left\lVert \epsilon^{\mathrm{T}}\right\rVert \left\lVert P_b\right\rVert   \left\lVert  \dot{\hat{\Pi}}_b
  \left(\sum_{r \in \mathcal{L}}(\Psi_r \otimes I_p )\right)^{-1}\sum_{k \in \mathcal{L}}(\Psi_k \otimes I_p ) \bar{\zeta}_k\right\rVert \\
  \leq& \left\lVert \epsilon\right\rVert^2 + \left\lVert P_b\right\rVert  ^2 \left\lVert \dot{\hat{\Pi}}_b
  \left(\sum_{r \in \mathcal{L}}(\Psi_r \otimes I_p )\right)^{-1}\sum_{k \in \mathcal{L}}(\Psi_k \otimes I_p ) \bar{\zeta}_k\right\rVert^2 \\
  \leq& \left(\frac{1}{4} \sigma_{\rm min}(Q_b)\! - \frac{1}{2}b_{ 1 } \! \right)\! \left\lVert \epsilon\right\rVert ^2 \!+\! \frac{\left\lVert P\right\rVert  ^2}{ \left(\frac{1}{4} \sigma_{\rm min}(Q_b) - \frac{1}{2}b_{ 1 } \! \right)} b_{\Pi}^2 {\rm e}^{-2 \beta_{\Pi}t}\\
  \leq&
  \left(\frac{1}{4} \sigma_{\rm min}(Q_b) - \frac{1}{2}b_{ 1 } \right)\left\lVert \epsilon\right\rVert ^2 +b_{21}{\rm e}^{-2\beta_{21}t}.
\end{aligned}
\end{equation}
By Lemma \ref{Lemma 6}, we conclude that $\dot{\hat{\Pi}}$ converges to zero exponentially. Since $\tilde{\zeta}$ also converges exponentially to zero, we get
\begin{equation}\label{EQ62}
\begin{aligned}
   & 2\left\lVert \epsilon^{\mathrm{T}}\right\rVert \left\lVert P\right\rVert  \left\lVert \dot{\hat{\Pi}}_b+  \mu_2 \hat{\Pi}_b G_b \sum_{k \in \mathcal{L}}(\Psi_k \otimes I_p )  \right\rVert  \left\lVert \tilde{\zeta}\right\rVert \\
  \leq& \left(\frac{1}{4} \sigma_{\rm min}(Q_b) - \frac{1}{2}b_{ 1 } \right)\left\lVert \epsilon\right\rVert ^2+b_{22} {\rm e}^{-2\beta_{22}t}.
\end{aligned}
\end{equation}
Next,
%\begin{equation}
%\begin{aligned}
\begin{flalign}
\epsilon_i^{\mathrm{T}} P_i  B_i\tilde{\chi}_i
        =& \epsilon_i^{\mathrm{T}} P_i  B_i \chi_i -\frac{\left\lVert \epsilon_i^{\mathrm{T}} P_i B_i \right\rVert^2}{ \left\lVert \epsilon_i^{\mathrm{T}} P_i B_i \right\rVert +\omega} \hat{\rho_i} &\notag\\
   \leq&  \left\lVert \epsilon_i^{\mathrm{T}} P_i  B_i\right\rVert \left\lVert \chi_i \right\rVert - \frac{\left\lVert \epsilon_i^{\mathrm{T}} P_i B_i \right\rVert^2  }{\left\lVert \epsilon_i^{\mathrm{T}} P_i B_i \right\rVert + \omega} \hat{\rho_i} &\notag \\
=& \frac{\left\lVert \epsilon_i^{\mathrm{T}} P_i B_i \right\rVert ^2 \left(  \left\lVert \chi_i \right\rVert-  \hat{\rho}_i \right)+ \left\lVert \epsilon_i^{\mathrm{T}} P_i B_i \right\rVert\left\lVert \chi_i \right\rVert \omega}{\left\lVert \epsilon_i^{\mathrm{T}} P_i B_i \right\rVert +\omega } &\notag \\
=&\frac{\left\lVert \epsilon_i^{\mathrm{T}} P_i B_i \right\rVert ^2 \left(\frac{\left\lVert \epsilon_i^{\mathrm{T}} P_i B_i \right\rVert +\omega}{\left\lVert \epsilon_i^{\mathrm{T}} P_i B_i \right\rVert}||\chi_i||-\hat{\rho}_i \right)}{\left\lVert \epsilon_i^{\mathrm{T}} P_i B_i \right\rVert +\omega}.
\end{flalign}
%\end{aligned}
%\end{equation}
Note that $\mathrm{d}\left\lVert \chi_i \right\rVert/\rm{d} t$ is bounded in view of Assumption \ref{assumption 5}. Thus, if $\left\lVert \epsilon_i^{\mathrm{T}} P_i B_i \right\rVert\geq \bar{d} \geq \frac{\rm{d}||\chi_i||}{\rm{d} t}$, that is, $\frac{\bar{d}+\omega}{\bar{d}} \frac{d||\chi_i||}{\rm{d}t} -\dot{\hat{\rho}} \leq \bar{d}+\omega-\dot{\hat{\rho}} \leq -\omega < 0$, there exists $t_{\chi} > 0$ such that for all $t \geq t_{\chi}$, we have
\begin{equation}\label{EQ76}
     \left(\frac{\left\lVert \epsilon_i^{\mathrm{T}} P_i B_i \right\rVert +\omega}{\left\lVert \epsilon_i^{\mathrm{T}} P_i B_i \right\rVert}||\chi_i||-\hat{\rho}_i \right) \leq \left(\frac{\bar{d}+\omega}{\bar{d}}||\chi_i||-\hat{\rho}_i \right) <0.
\end{equation}
Thus, $\epsilon_i^{\mathrm{T}} P_i  B_i\tilde{\chi}_i <0$ and $\epsilon^{\mathrm{T}} P_b  B_b\tilde{\chi}<0$ over $t \in[t_{\chi},\infty)$.

According to (\ref{EQ81}) $\sim$ (\ref{EQ76}), we have
\begin{equation}\label{EQ70}
  \dot{V} \leq -\left(b_{ 1 }+\frac{1}{2} \sigma_{\rm min}(Q_b) \right) \epsilon^{\mathrm{T}} \epsilon +b_{ 2 }{\rm e}^{-\beta_{2}t}, ~t \geq t_{\chi}.
\end{equation}

Solving (\ref{EQ70}) yields
\begin{flalign}
&V(t) \leq  V(0) \! -\! \!\int_{0}^{t} \! \left(b_{ 1 } \!+\!\frac{1}{2} \sigma_{\rm min}(Q_b) \right) \epsilon^{\mathrm{T}}\epsilon  \!\,{\rm d}\tau \!+\!\! \int_{0}^{t} \! b_{ 2 }{\rm e}^{-\beta_{2} \tau}\! \,{\rm d}\tau.&
\end{flalign}
Then, we obtain
\begin{equation}\label{EQ72}
    \epsilon^{\mathrm{T}} \epsilon \leq -\int_{0}^{t} \frac{1}{\sigma_{\rm min}(P)} b_{ 3 }  \epsilon^{\mathrm{T}}\epsilon\,{\rm d}\tau + \bar{B}.
\end{equation}
where $b_{ 3 } = b_{ 1 }+\frac{1}{2} \sigma_{\rm min}(Q_b) $ and $ \bar{B}=V(0)-\int_{0}^{t} b_{ 2 }{\rm e}^{-\beta_{2}\tau}\,{\rm d}\tau $ are bounded constants.
Recalling the Bellman-Gronwall Lemma, (\ref{EQ72}) is rewritten as
\begin{equation} \label{EQ 77}
  \left\lVert \epsilon \right\rVert \leq \sqrt{\bar{B}} {\rm e}^{-\frac{b_{ 3 } t}{2\sigma_{\rm min}(P_b)} }.
\end{equation}
According to (\ref{EQ76}) and (\ref{EQ 77}), we conclude that $\epsilon$ is bounded by $\bar{\epsilon}$, where $\bar{\epsilon}=[\bar{\epsilon}_1^{\mathrm{T}}, \bar{\epsilon}_2^{\mathrm{T}}, \dots,\bar{\epsilon}_N^{\mathrm{T}}]^{\mathrm{T}}$ with
$||\bar{\epsilon}_i||=\frac{\bar{d}}{\sigma_{\rm min}(P_i B_i)}$ for $i= \textbf{I}[1,N]$.

Hence, according to Lemma \ref{Lemma5}, the global output synchronization error satisfies
\begin{equation}
\begin{aligned}
e =& y - \left(\sum_{r\in \mathcal{L} }(\Phi_r \otimes I_p)\right)^{-1} \sum_{k \in \mathcal{L} } (\Psi_k \otimes I_p) \underline{y}_k \\
=&y-(I_N \otimes R)(\zeta-\tilde{\zeta})\\
=& {\rm blkdiag}(C_i)x   -{\rm blkdiag}(C_i \hat{\Pi}_i)\zeta +{\rm blkdiag}(C_i \hat{\Pi}_i)\zeta\\
&-(I_N \otimes R)\zeta +(I_N \otimes R) \tilde{\zeta} \\
=&  {\rm blkdiag}(C_i)\epsilon  -{\rm blkdiag}(\tilde{R}_i)\zeta +(I_N \otimes R) \tilde{\zeta}.
\end{aligned}
\end{equation}

%{\color{black}
Since we proved that $\epsilon$, $\tilde{R}_i$, and $\tilde{\zeta}$ converge to 0 exponentially, the global output containment error $e$ is bounded by $\bar{e}$, with $\bar{e}=[\bar{e}_1^{\mathrm{T}}, \bar{e}_2^{\mathrm{T}}, \dots ,\bar{e}_i^{\mathrm{T}}]^{\mathrm{T}}$ and $\bar{e}_i=\frac{\bar{d} ||C_i||}{\sigma_{\rm min}(P_i B_i)}$ for $i= \textbf{I}[1,N]$.
%}
Thus, the proof is completed.
$\hfill \hfill \blacksquare $

\begin{remark}
Compared with \cite{zuo2020}, which only solved the containment problem against unbounded attacks, we consider a more challenging output
containment problem with multiple leaders against composite attacks. The boundary of the output containment error is explicitly given by $\frac{ ||C_i|| \bar{d}}{\sigma_{\rm min}(P_i B_i)}$.
%It's helpful for designers to pre-estimate of the controller.
$\hfill \hfill  \square $
\end{remark}

\section{Examples}%\label{SecSm}

In this section, we present two examples to demonstrate effectiveness of the proposed control approach.

\subsection{Example 1}
We consider a MAS consisting of seven agents (three leaders indexed by 5$\sim$7 and four followers indexed by 1$\sim$4), with the corresponding graph shown in Fig. \ref{fig:figure3}. Set $a_{ij}=1$, if there exists a path from node $j$ to node $i$.
Let the leader dynamics be described by
$$
S
=
 \left[
  \begin{array}{cc}
 0.5  & -0.4  \\
 0.8 &  0.5   \\
  \end{array}
  \right],R
  =
   \left[
    \begin{array}{cc}
   1 &  0    \\
   0  & 1  \\
    \end{array}
    \right],
$$
and the follower dynamics be given by
$$
A_1
\!=\!
 \left[\!
  \begin{array}{cc}
 3  & -2  \\
 1 &  -2   \\
  \end{array}
 \! \right],B_1
 \! =\!
   \left[\!
    \begin{array}{cc}
   1.8 &  -1    \\
   2  & 3  \\
    \end{array}
   \! \right],C_1
  \!=\!
   \left[\!
    \begin{array}{cc}
   -0.5 &  1    \\
  2  & -1.5  \\
    \end{array}
   \! \right],
$$
$$
A_2
\!=\!
 \left[\!
  \begin{array}{cc}
 0.6  & -1  \\
 1 &  -2   \\
  \end{array}
  \!\right],B_2
  \!=\!
   \left[\!
    \begin{array}{cc}
   1 &  -2    \\
   1.9  & 4  \\
    \end{array}
   \! \right],C_2
  \!=\!
   \left[\!
    \begin{array}{cc}
   -0.5 &  1    \\
  1.5  & 1.4  \\
    \end{array}
  \!  \right],
$$
$
A_3=A_4
=
 \left[
  \begin{array}{ccc}
 0  &  1  &0\\
 0 &  0 & 1  \\
  0 &  0 & -2  \\
  \end{array}
  \right],B_3
  =B_4=
   \left[
    \begin{array}{ccc}
   6 &  0    \\
   0  & 1  \\
   1  & 0 \\
    \end{array}
    \right],$

    \noindent
    $
    C_3=C_4
  =
   \left[
    \begin{array}{ccc}
   0.5 &  -0.5  & 0.5  \\
  -0.5  & -0.5  &0.5\\
    \end{array}
    \right].
$

The DoS attack periods are assigned as $[0.5+2k, 1.53+2k)s$ for $k\in \mathbb{N}$, which satisfy Assumption 3. The MAS can defend against FDI attacks and camouflage attacks, since the information transmitted on the CPL is not used in the hierarchical control scheme.
The actuation attack signals are designed as $\chi_1=\chi_2=\chi_3=0.01 \times [2t \quad t]^{\mathrm{T}}$ and $\chi_4=-0.01\times[2t\quad t]^{\mathrm{T}}$.

 %the   light red zone denotes the UUB bound || eTi PiBi|| = d

The gains in (\ref{EQ15}), (\ref{equation 200}), and (\ref{EQ48}) are set to $\mu_1=2$, $\mu_2=0.5$, and $\mu_3=6$, respectively. According to Theorem 2, $\bar{P}_2 =\left[ \begin{array}{cc}
    0.6038 & 0.0108 \\
    0.0108 & 0.1455
\end{array}\right]$ and $G=\left[\begin{array}{cc}
     4.5191& 0.5186 \\
     0.5186& 4.0746
\end{array}\right]$. By employing the above parameters, the estimated trajectories of the leader dynamics are shown in Fig. \ref{fig:figure4}. The estimated leader dynamics $||\hat{\Upsilon}_i||$ under DoS attacks remains stable after $6 s$, and the stable value corresponds to the true leader dynamics. The state estimation errors on the TL are shown in Fig. \ref{fig:figure5}, demonstrating that the performance of the TL is reliable under DoS attacks.
The trajectories of the solution errors of the output regulator equations are shown in Fig. \ref{fig:figure6}, which are consistent with Theorem 3.
To analyze the actuation attacks on the CPL, we design the controller gains (\ref{EQ0968}) as
$
K_1
=
 \left[
  \begin{array}{cc}
 -1.71  & 0.07  \\
 2.31  &  -1.12   \\
  \end{array}
  \right],K_2
  =
   \left[
    \begin{array}{cc}
   -0.62 &  -0.29   \\
   1.16  & -0.56  \\
    \end{array}
    \right],$
    $K_3=K_4
  =
   \left[
    \begin{array}{ccc}
   -1.00 &  -0.19  & -0.08  \\
  0.02  & -0.96  &-0.30\\
    \end{array}
    \right].
$

Then, we solve \textbf{Problem ACMCA} according to Theorem 4. Fig. \ref{fig:figure7} displays the output trajectories of all agents. The followers eventually enter the convex hull formed by the leaders. The output containment errors of the followers are shown in Fig. \ref{fig:figure8}, demonstrating that the local output error $e_i(t)$ is UUB. The tracking error between the CPL and TL is displayed in Fig. \ref{fig:figure9}, showing that the tracking error $\epsilon_i$ is UUB with the prescribed error bound. The obtained results demonstrate that the output containment problem can be reliably solved even under composite attacks.

\subsection{Example 2}

%\begin{figure}[!]
%  %\begin{minipage}[t]{1\linewidth}
%  \centering
%  \includegraphics[width=0.45\textwidth]{picnew2/tp23.pdf}
%  \caption{Communication topology graph of a UAV swarm, with blue and gray UAVs representing leaders and followers, respectively.}
%  \label{fig:figure10}
%\end{figure}

Next, we consider a practical example involving unmanned aerial vehicles (UAVs), whose corresponding graph is shown in Fig. \ref{fig:figure10}. We assume that all edges have a weight of 1.

%  Choose
%  $$\mathcal{A}_f=\left[ \begin{array}{ccccc}
%      0 & 0& 0& 0& 1 \\
%       1 & 0& 0& 0& 1\\
%       1 & 1& 0& 0& 0\\
%       0 & 0& 1& 0& 0\\
%       0 & 0& 0& 1& 0\\
%  \end{array}
%  \right],
%  $$
%  and
%   $$G_{ik}=\left[ \begin{array}{ccccc}
%      1 & 0& 0\\
%       0 & 0&0 \\
%       0 & 0&0\\
%       0 & 0&1 \\
%       0 & 1& 0\\
%  \end{array}
%  \right],
%  $$
%  where $\mathcal{A}_f$ represent  the associated adjacency matrix between followers and $G_{ik}={\rm diag}(g_{ik})$ with $g_{ik}$  is the weight of the path from $i$th leader to $k$th follower.

 The dynamics of the leader and follower UAVs can be approximated by the following second-order systems \cite{wang2018optimal,dong2016time}:
 \begin{equation}
  \begin{cases}
 \dot{p}_k(t)=v_k(t), \\
 \dot{v}_k(t)=\alpha_{p_0} p_k (t) + \alpha_{v_0} v_k(t),\\
 y_k(t)=p_k(t),
\end{cases}
 \end{equation}
and
\begin{equation}
  \begin{cases}
 \dot{p}_i(t)=v_i(t), \\
 \dot{v}_i(t)=\alpha_{p_i} p(t) + \alpha_{v_i} v_i(t)+u_i(t),\\
 y_i(t)=p_i(t),
\end{cases}
\end{equation}
where $p_k(t)$ and $v_k(t)$ ($p_i(t)$ and $v_i(t)$) represent the position and velocity of the $k$th leader UAV ($i$th follower UAV), respectively. The damping constants of the leaders are set to $\alpha_{p_0}=-0.6$ and $\alpha_{v_0}=0$. The damping constants of the followers are assigned as $\alpha_{p_1}=-1, \alpha_{v_1}=-1$, $\alpha_{p_2}=-0.5, \alpha_{v_2}=-1.2$, $\alpha_{p_3}=-1.2, \alpha_{v_3}=-1$, $\alpha_{p_4}=-0.8, \alpha_{v_4}=-0.5$, and $\alpha_{p_5}=-0.4, \alpha_{v_5}=-1.2$. The gains are set to $\mu_1=5$, $\mu_2=1$, and $\mu_3=7$. We assign the DoS attack period as $[0.2+2k, 1.86+2k)s$, for $k\in \mathbb{N}$, and the actuation attack signals are defined as $\chi_1=[0.01t\quad0.02t\quad0.02t]^{\mathrm{T}},\chi_2=[0.02t\quad0.01t\quad0.02t]^{\mathrm{T}},\chi_3=[0.02t\quad0.02t\quad0.02t]^{\mathrm{T}},\chi_4=[-0.01t\quad-0.02t\quad-0.02t]^{\mathrm{T}},\chi_5=[-0.02t\quad-0.01t\quad0.02t]^{\mathrm{T}}$.

%\begin{figure}[!]
%  %\begin{minipage}[t]{1\linewidth}
%  \centering
%  \includegraphics[width=0.45\textwidth]{picnew/Upsilon.eps}
%  \caption{Leader dynamics on the TL in Theorem 1: Shadowed areas denote time intervals under DoS attacks.}
%  \label{fig:figure11}
%\end{figure}
%
%
%  %\begin{figure*}[htbp]
%\begin{figure}[htbp]
%  \centering
%    \includegraphics[width=0.45\textwidth]{picnew/zeta31026.eps}
%  \caption{Distributed estimation performance of the TL: Shadowed areas denote time intervals under DoS attacks.}
%   \label{fig:figure12}
%\end{figure}

{Set the initial values of the leader dynamics observers (\ref{EQ15}) and the output regulator equation solvers (\ref{EQ48}) to zero.} Then, we assign the initial leader states as $p_6=[1\quad1 \quad 1]^{\mathrm{T}}, v_6=[4 \quad 5\quad4]^{\mathrm{T}}$, $p_7=[4 \quad 1 \quad 2]^{\mathrm{T}}, v_7=[5 \quad 6 \quad 6]^{\mathrm{T}}$, and $p_8=[6 \quad 5 \quad 8]^{\mathrm{T}}, v_6=[4 \quad 6 \quad 5]^{\mathrm{T}}$ and randomly set remaining initial values. According to Theorems 2 and 4, the state estimator gain of the TL is calculated as $G=\left[\begin{array}{cc}
     2.8483 & 0.2594 \\
     0.2594 & 3.0611
 \end{array}\right]$, and the gains of the CPL are given by $K_1=[-0.4142 \quad 	-0.6818]$, $K_2= [-0.6180 \quad 	-0.7173]$, $K_3=[-0.3620 \quad 	-0.6505]$, $K_4=[-0.4806	 \quad -0.9870]$, $K_5= [-0.6770	 \quad -0.7478]$.

 %\begin{figure}[!]
%  %\begin{minipage}[t]{1\linewidth}
%  \centering
%  \includegraphics[width=0.45\textwidth]{picnew/Delta.eps}
%  \caption{Estimated solutions of the output regulator equations in Theorem 3: Shadowed areas denote time intervals under DoS attacks.}
%  \label{fig:figure13}
%\end{figure}
%
%
%\begin{figure}[!]
%  %\begin{minipage}[t]{1\linewidth}
%  \centering
%  \includegraphics[width=0.45\textwidth]{picnew/track/track3.eps}
%  \caption{Trajectories of the UAV swarm between $0 \sim 30 s$. Hollow and filled squares represent the start and end points of the UAVs, respectively.}
%  \label{fig:figure14}
%\end{figure}
%
%
%
%
%    \begin{figure}[htbp]
%  \centering
%  \includegraphics[width=0.45\textwidth]{picnew/ex.eps}
%  \caption{Output containment performance of the proposed control scheme.}
%  \label{fig:figure15}
%  \end{figure}
%
%
%
%\begin{figure}[htbp]
%  %\begin{minipage}[t]{1\linewidth}
%  \centering
%  \includegraphics[width=0.45\textwidth]{picnew/ebar.eps}
%  \caption{Estimation errors between the TL and CPL: The blue shadow area indicates the UUB bound of $\bar{d}$.}
%  \label{fig:figure16}
%\end{figure}

Fig. \ref{fig:figure11} reveals that $\hat{\Upsilon}_i$ converges to $\Upsilon$ under DoS attacks, which implies that the modeling errors of the leader dynamics converge to zero. The state estimator errors of the TL are shown in Fig. \ref{fig:figure12}, demonstrating reliable performance of the TL under DoS attacks. Fig. \ref{fig:figure13} verifies validity of Theorem 3.
The output trajectories of the reference UAVs are displayed over time in Fig. \ref{fig:figure14}, where the initial and final positions of the UAVs are marked by hollow and filled squares, respectively. The trajectory of each follower remains in a small neighborhood around the dynamic convex hull spanned by the leaders after $t=10 s$, that is, output containment is achieved, which can also be seen in Fig. \ref{fig:figure15}.
Fig. \ref{fig:figure16} verifies validity of the upper bound of the output containment errors.

% \begin{figure}[!htbp]
% %\begin{minipage}[t]{1\linewidth}
% \centering
% \includegraphics[width=0.6\textwidth]{4Ag.pdf}
% \caption{Time-varying directed communication topology among all agents}
% \label{fig:figure1}
% \end{figure}

%{\color{blue}
%\begin{figure}[htbp]
%\centering
%\subfigure[Performance of observer w.r.t. the leader]{
%\begin{minipage}[t]{0.475\textwidth}
%\centering
%\includegraphics[width=0.85\textwidth]{pic/pobs.eps}
%%\caption{fig1}
%\end{minipage}\label{fig:figure2:1}
%}
%%\hspace{-0.1in}
%\subfigure[Performance of observer w.r.t. the first leader]{
%\begin{minipage}[t]{0.475\textwidth}
%\centering
%\includegraphics[width=0.85\textwidth]{pic/vobs.eps}
%%\caption{fig2}
%\end{minipage}\label{fig:figure2:2}
%}\\%
%\centering
%\caption{Performance of two observers}
%\label{fig:figure2}
%\end{figure}

\section{Conclusions}
The distributed resilient output containment problem for heterogeneous MASs against composite attacks is studied in this paper. Inspired by hierarchical protocols, a TL that can resist to most attacks (including FDI attacks, camouflage attacks, and actuation attacks) is designed to decouple the defense strategy into two tasks. The first one considers defense against DoS attacks on the TL, and the other one considers defense against unbounded actuation attacks on the CPL.
To address the first task, we introduce a TL to correct modeling errors of the leader dynamics and defend against DoS attacks. Then,
distributed observers and estimators are used to reconstruct the leader dynamics and the follower states under DoS attacks on the TL. To address the second task, output regulator equation solvers and adaptive decentralized control schemes are introduced to address the output containment problem and resist unbounded actuation attacks on the CPL. Finally, we prove that the
%Editor: Please ensure that the intended meaning has been maintained in the following edit.
upper bound
of the output containment error is UUB under composite attacks and calculate the error bound explicitly. Simulations are provided to demonstrate effectiveness of the proposed methods.

\

% Proof: Consider the Lyapunov function candidate
% $$
% V_{1}=\frac{1}{2} \sum_{i=1}^{N} \xi_{i}^{T} P \xi_{i}+\sum_{i=1}^{N} \sum_{j=1, j \neq i}^{N} \frac{\left(c_{i j}-\alpha\right)^{2}}{8 \kappa_{i j}}
% $$
% where $\alpha$ is a positive constant that is to be determined later. Evidently, $V_{1}$ is positive definite. The time derivative of $V_{1}$ along the trajectory of (5) is given by
% $$
% \begin{aligned}
% \dot{V}_{1}=& \sum_{i=1}^{N} \xi_{i}^{T} P \dot{\xi}_{i}+\sum_{i=1}^{N} \sum_{j=1, j \neq i}^{N} \frac{c_{i j}-\alpha}{4 \kappa_{i j}} \dot{c}_{i j} \\
% =& \sum_{i=1}^{N} \xi_{i}^{T} P A \xi_{i}+\sum_{i=1}^{N} \xi_{i}^{T} P B K \sum_{j=1}^{N} c_{i j} a_{i j}\left(\tilde{x}_{i}-\tilde{x}_{j}\right) \\
% &+\sum_{i=1}^{N} \sum_{j=1, j \neq i}^{N} \frac{c_{i j}-\alpha}{4 \kappa_{i j}} \dot{c}_{i j}
% \end{aligned}
% $$
% Since $a_{i j}=a_{j i}$ and $c_{i j}(t)=c_{j i}(t)$, it can be easily verified that
% $$
% \begin{array}{rl}
% \sum_{i=1}^{N} \xi_{i}^{T} & P B K \sum_{j=1}^{N} c_{i j} a_{i j}\left(\tilde{x}_{i}-\tilde{x}_{j}\right) \\
% =&-\frac{1}{2} \sum_{i=1}^{N} \sum_{j=1}^{N} c_{i j} a_{i j}\left(\xi_{i}-\xi_{j}\right)^{T} \Gamma\left(\tilde{x}_{i}-\tilde{x}_{j}\right)
% \end{array}
% $$

\bibliography{PIDFR}
\end{document}